\documentclass[aps,prb,reprint,footinbib]{revtex4-2}

\pdfoutput=1
\usepackage{graphicx}
\usepackage{amsmath,amsfonts,amssymb,amsthm,mathtools,mathrsfs}
\usepackage{physics}
\usepackage{hyperref}
\usepackage{booktabs}
\usepackage{url}
\usepackage{leftidx}
\usepackage{gensymb}
\usepackage{textcomp}
\usepackage{lmodern}
\usepackage{array}
\usepackage{comment}
\usepackage[dvipsnames]{xcolor} 
\usepackage{placeins} 
\usepackage[danish,english]{babel}
\usepackage{dsfont}

\definecolor{cerulean}{rgb}{0.0, 0.48, 0.65}
\definecolor{emerald}{rgb}{0.31, 0.78, 0.47}
\definecolor{blue(ncs)}{rgb}{0.0, 0.53, 0.74}

\hypersetup{
     colorlinks=true,
     linkcolor=magenta,
     filecolor=blue,
     citecolor=emerald,      
     urlcolor =blue(ncs),
}

\DeclareMathAlphabet{\pazocal}{OMS}{zplm}{m}{n}

\addto\captionsenglish{}

\usepackage{tikz}
\usetikzlibrary{patterns}
\usetikzlibrary{automata,positioning}
\usetikzlibrary{calc}
\usepackage{tikz-3dplot}
\usetikzlibrary{arrows.meta}
\usepackage[compat=1.1.0]{tikz-feynman} 
\usepackage{contour}
\usetikzlibrary{shapes.geometric}

\newcommand{\overbow}[1]{
\,
   \tikz [baseline = (N.base), every node/.style={}] {
      \node [inner sep = 0pt] (N) {$#1$};
      \draw [line width = 0.4pt] plot [smooth, tension=1.3] coordinates {
         ($(N.north west) + (-0.2,-0.2)$)
         ($(N.north)      + (0,0.6ex)$)
         ($(N.north east) + (0.2,-0.2)$)
      };
   }
   \,
}

\newcommand{\overarrow}[1]{
\,
   \tikz [baseline = (N.base), every node/.style={}] {
      \node [inner sep = 0pt] (N) {$#1$};
      \draw [->,line width = 0.4pt] plot [smooth, tension=1.3] coordinates {
         ($(N.north west) + (-0.2,-0.2)$)
         ($(N.north)      + (0,0.6ex)$)
         ($(N.north east) + (0.2,-0.2)$)
      };
   }
   \,
}

\begin{document}

\title{On the Ferrimagnetic State of \texorpdfstring{CrCl$_2$(pyz)$_2$}{CrCl2(pyz)2}}

\author{Freja Schou Guttesen}
\email{freja.guttesen@nbi.ku.dk}
\affiliation{Niels Bohr Institute, University of Copenhagen, Jagtvej 155A, DK-2200 Copenhagen, Denmark}
\affiliation{Sino-Danish College (SDC), University of Chinese Academy of Sciences}

\author{Per Hedegård}
\email{hedegard@nbi.ku.dk}
\affiliation{Niels Bohr Institute, University of Copenhagen, Jagtvej 155A, DK-2200 Copenhagen, Denmark}

\date{\today}

\begin{abstract}
Van der Waals layered ferromagnetic compounds with high two-dimensional electronic conductivity holds strong potential for quantum computing, future unconventional superconductors, catalysts, batteries, and fuel cells. 
We suggest a minimal theoretical model to understand the magnetic properties of the metal-organic framework CrCl$_2$(pyz)$_2$ (pyz=pyrazine). Using a Hubbard model we show that the groundstate is dominated by a specific configuration of delocalized electrons on the pyz sites with a ferrimagnetic coupling to the localized spins on the Cr sites. This model suggests a magnetic moment of $2\mu_B$ which is remarkably close to the experimental value of $1.8 \mu_B$ [K. S. Pedersen et al., Nat. Chem. 10, 1056-1061 (2018)]. From Weiss mean-field theory we predict a weak ferromagnetic Cr-Cr coupling of $\approx 0.9$ meV. This is consolidated by second order perturbation theory of the RKKY interaction yielding a Cr-Cr coupling of $\approx 5$ meV. Understanding the interactions in these types of compounds can facilitate designs of metal-organic compounds with tailored magnetic properties.
\end{abstract}

\maketitle
\FloatBarrier
\section{Introduction}
\FloatBarrier

Metal-organic frameworks (MOFs) have been predicted to serve as a platform for the next generation of quantum technology~\cite{Low-Dim, guilty}. The diversity of organic ligands compared to metal ligands allow for enhanced tunability of electronic and magnetic properties. These compounds may be key to harvest renewed insights into and exploitation of quantum phenomena such as the Quantum Hall Effect~\cite{OGQH, QuantumHall, quantumHall2}, topological insulators~\cite{topinsulators} and superconductivity~\cite{super}. 
The putatively first organic-inorganic hybrid 2D MOF, CrCl$_2$(pyz)$_2$ (pyz = pyrazine), was successfully synthesized in 2018~\cite{Pedersenetal}.
CrCl$_2$(pyz)$_2$ exhibits both long-range magnetic order and high 2D electronic conductivity. The compound has a Curie temperature of 55 K, and its magnetization at low temperatures saturates to 1.8 $\mu_B$ in strong external fields. This indicates an intralayer ferrimagnetic coupling in agreement with predictions by density functional theory (DFT)~\cite{ChemC, Xie_2020}. First principle calculations indicate that CrCl$_2$(pyz)$_2$, like other hybrid materials~\cite{synthesis, RisingUp, MetalOrganicMagnets}, can be exfoliated making it relevant for 2D applications~\cite{ChemC}. The majority of 2D materials are non-magnetic in their pristine form~\cite{2dnonmag}. A magnetic 2D material, like CrCl$_2$(pyz)$_2$, could have applications within spintronics~\cite{spintronices}, multiferroics~\cite{multi, Fiebig2016, Spaldin2019} and magnetoelectronics~\cite{magneto}. This material could possibly even have applications within dark matter detection~\cite{DM1, DM2, DM3}. 
The sibling coordination solids VCl$_2$(pyz)$_2$ and TiCl$_2$(pyz)$_2$ display contrasting physical properties. In the former, divalent V ions facilitate an antiferromagnetically ordered ground state, while in the latter trivalent Ti ions favor paramagnetism and a correlated Fermi liquid state~\cite{PerlepeEA22}. In principle, this opens the possibility for tuning through exotic phase transitions via doping. This may be compared to the magnetic phase transitions proposed for Mn$_{2-x}$Cr$_2$Sb~\cite{JarrettEA64,magnetotran} and Mn$_2$Sb$_{1-x}$Bi$_x$~\cite{ShenEA21}. Finally, the perovskite S$_2$FeMoO$_6$ may be highlighted as another material in which the coexistence of itinerant and localized electrons cause an ordered, in this case putatively ferromagnetic, ground state~\cite{GarciaLandaEA99, TovarEA02}.

A crucial step towards synthesizing materials with desired magnetic properties is to understand the mechanism yielding these properties. 
Here we provide a simple explanation of the ferrimagnetic coupling and explain why the magnetic moment of CrCl$_2$(pyz)$_2$ is approximately 2$\mu_B$. As such, we propose a minimal model for monolayer CrCl$_2$(pyz)$_2$ that captures the essential features of chromium couplings mediated by intermediate and itinerant pyrazine electrons. 

First, we compute a tight-binding model using a Slater--Koster procedure to obtain hopping parameters. From this we propose an effective model relevant for the valence electrons. The ground state configuration of the model is investigated before a magnetic field is introduced and the magnetic moment is computed. Using mean-field theory an estimate of the Cr-Cr coupling is found. Finally, the exchange interaction between electrons on neighboring chromium and pyrazine sites is calculated and an expression of the indirect exchange interaction coupling between neighboring chromium sites is obtained using second order perturbation of the RKKY interaction. From this we predict an estimate of the coupling between the localized Cr spins and the itinerant pyz electrons.
\FloatBarrier
\section{Tight Binding Model of Monolayer $\textbf{CrCl}_\textbf{2}\textbf{(pyz)}_2$}
\begin{figure}[t!hb]
\centering
\includegraphics[width=\linewidth]{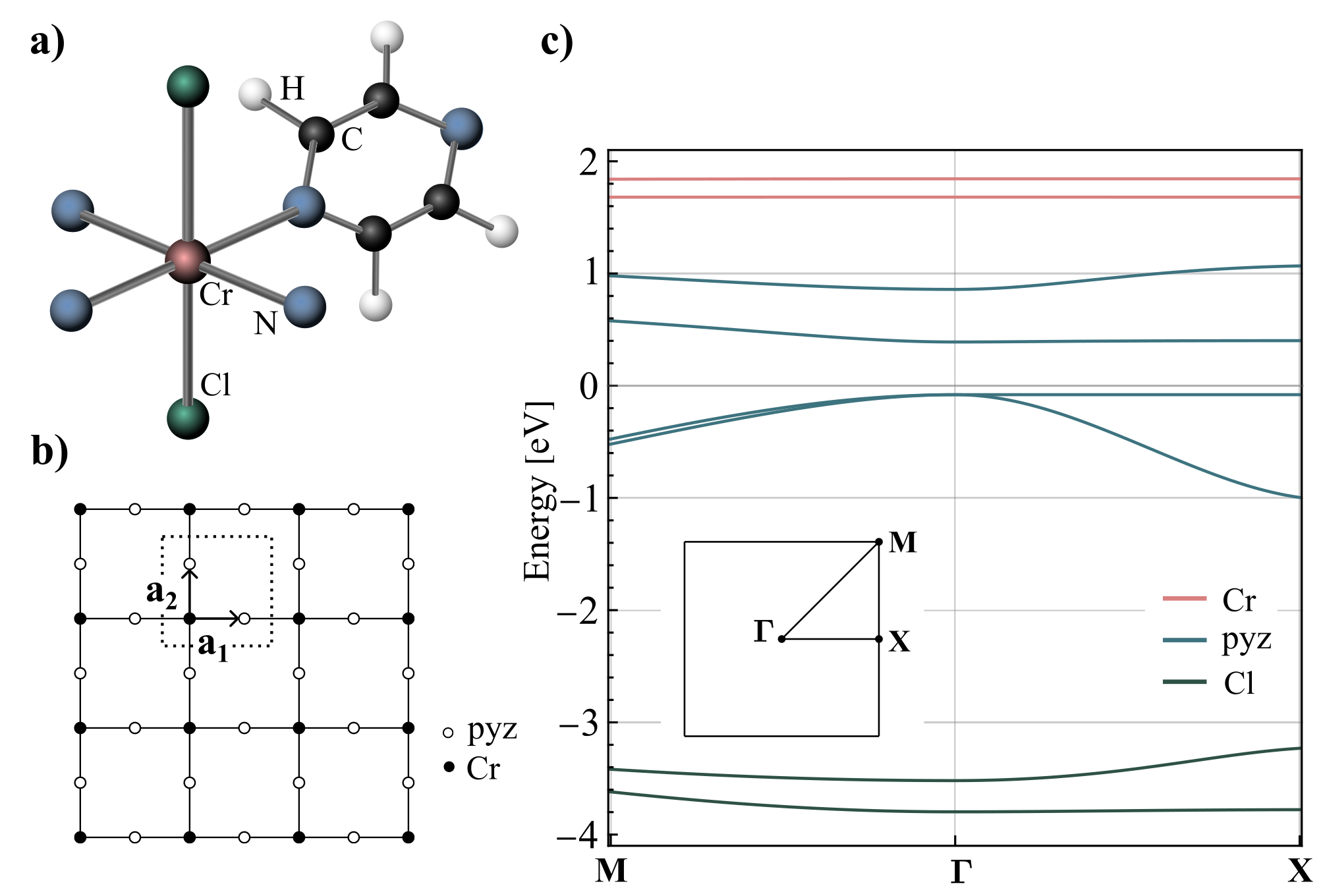}
\caption{\textbf{a)} Illustration of a fragment of one unit cell of CrCl$_2$(pyz)$_2$. \textbf{b)} Monolayer CrCl$_2$(pyz)$_2$ viewed along the Cl-Cr-Cl axis. The chromium and pyrazine sites are illustrated with black and white circles, respectively. Each pyz ring has been approximated by a single sites. Further, the unit cell is marked with a dashed square and the relevant lattice vectors, $\mathbf{a_1}$ and $\mathbf{a_2}$, are indicated. \textbf{c)} Band structure of monolayer CrCl$_2$(pyz)$_2$ obtained from tight binding along the path $\mathbf{M} \rightarrow \mathbf{\Gamma} \rightarrow \mathbf{X}$. Inset show the Brillouin zone and the relevant symmetry points.}
\label{fig:goodstuff}
\end{figure}

\FloatBarrier
Monolayer CrCl$_2$(pyz)$_2$ has an approximately four-fold rotational in-plane symmetry. Each layer is coupled through Van der Waals interactions. The pyrazine rings are tilted with two alternating orientations~\cite{Pedersenetal}, see FIG.~\ref{fig:onemolCrCl} in the Supplementary Material. One unit cell contains two Cl atoms, one Cr atom, and two pyrazine rings, see a fragment of a unit cell in Fig.~\ref{fig:goodstuff} \textbf{a)}. 

We compute an effective tight-binding model for monolayer CrCl$_2$(pyz)$_2$, as summarized in FIG.~\ref{fig:goodstuff} \textbf{b)}. The lattice is approximated as a square lattice in the Cr-pyz plane, and each pyrazine ring is modeled by a single site with two orbitals tilted by $45^\circ$ compared to the in-plane Cr lattice, see FIG.~\ref{fig:onemolCrCl}. Due to the tilting of the pyrazine sites, the Cr sites and the pyz sites form a Lieb lattice~\cite{Lieb}. A study of the symmetries of the model complex CrCl$_2$(pyz)$_4$ using a group theoretical approach can be found in the Supplementary Material. The basis includes five d-orbitals on the Cr site, three p-orbitals on both Cl sites, and two p$_{z'}$ orbitals tilted in two different orientations on the two pyz sites, $\mathbf{\Phi}_\textbf{k}=\mathbf{\Phi}_\textbf{k}^{\text{Cr}}\otimes \mathbf{\Phi}_\textbf{k}^{\text{Cl}}\otimes\Phi_\textbf{k}^{\text{pyz}}$. Here $z'$ refers to the direction perpendicular to the plane spanned by the pyz rings. 
The Hamiltonian 
\begin{align}
H=\sum_\textbf{k}\mathbf{\Phi}_\mathbf{k}^\dagger \hat{\mathcal{H}}(\textbf{k})\mathbf{\Phi}_\textbf{k}^{\phantom\dagger}
\end{align}
describes inter- and intraorbital hopping between the aforementioned orbitals, as encoded by the $15\cross 15$ matrix $\hat{\mathcal{H}}$. The form of $\hat{\mathcal{H}}$ is listed in Eq.~\eqref{eq:Hk} in the Supplementary Material. The basis is not spin-resolved. The elements in the Hamiltonian has been obtained using the Slater--Koster decomposition of the tight binding model~\cite{Harrison, SlaterKoster}.\\
\\
The band structure along the path $\mathbf{M} \rightarrow \mathbf{\Gamma} \rightarrow \mathbf{X}$ is displayed in FIG.~\ref{fig:goodstuff} \textbf{c)}. 
The Fermi Level is located at $0$ eV. This captures the qualitative, semi-metallic properties in agreement with first principle calculations~\cite{Pedersenetal}. 
The two empty d-bands of Cr are located above 1.5 eV, whereas the filled d-bands are located below the p-bands (not visible in the plot). This is also seen from first principle calculations of the PDOS~\cite{Pedersenetal} and could indicate a strong $\pi$-d hybridization. The flatness of the d-bands indicates a localization of the d-electrons in accordance with the behavior of a Mott insulator.
The p-levels from the pyz sites are dispersive which indicates a delocalization of the p-electrons. The Cl levels are located far below the Fermi level and will be safely neglected in the minimal model. From first principle calculations the summed spin population on the pyrazine rings is calculated to $-0.68$ delocalized over four pyz sites of the model complex CrCl$_2$(pyz)$_4$~\cite{Pedersenetal}.
\FloatBarrier
\FloatBarrier
In this paper the two pyrazine rings in a unit cell will be approximated as two single sites over which one electron delocalize. Further, Ref.~\cite{Pedersenetal} concludes that the total spin on the Cr site is $S=3/2$. For these reasons, the Cr sites will be assumed to have a localized spin $S=3/2$, and we assume one spin $S=1/2$ to be delocalized over the two pyz sites.

\FloatBarrier
\section{Minimal Model}
\begin{figure}[b!th]
\centering
\includegraphics[width=.8\linewidth]{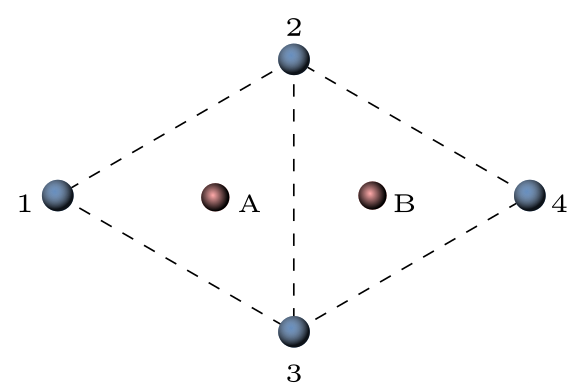}
\caption{Model containing two chromium sites (red circles labeled A and B) and four pyrazine sites (blue circles labeled 1-4). Each chromium spin couples only to three out of the four pyrazine sites (A to 1-3 and B to 2-4).}
\label{fig:simplemodel}
\end{figure}

\FloatBarrier
To shed light on the microscopic mechanism of the magnetic ground state of CrCl$_2$(pyz)$_2$ we investigate the model shown in FIG.~\ref{fig:simplemodel}. This is not identical to two unit cells of a single layer of CrCl$_2$(pyz)$_2$ since the pyz sites are moved with respect to the two chromium atoms. The model does, however, capture the competition between kinetic energy and exchange energy, which comprise the essential low-energy physics relevant for explaining the magnetic properties of CrCl$_2$(pyz)$_2$. In this model the two chromium atoms are assumed to be localized $S=3/2$ spins. There are two electrons that can delocalize over the four pyrazine sites. The Cl sites are omitted in this description, yet effectively taken into account by reducing the number of electrons on the Cr sites.
We include the exchange coupling between the chromium spins and the spins of the two pyrazine electrons. Moreover, the pyrazine electrons are allowed to hop between the pyrazine sites (indicated with dashed lines in FIG.~\ref{fig:simplemodel}). Lastly, we include an on-site repulsive Hubbard interaction, $U$, penalizing electrons residing on the same site. The effective Hamiltonian can thus be written as
\begin{align}
\hat{\mathcal{H}}= &-2J\hat{\vec{S}}_A\cdot\left(\hat{\vec{S}}_1+\hat{\vec{S}}_2+\hat{\vec{S}}_3\right)-2J\hat{\vec{S}}_B\cdot\left(\hat{\vec{S}}_2+\hat{\vec{S}}_3+\hat{\vec{S}}_4\right)\nonumber\\
&-t\sum\limits_{\sigma=\uparrow,\downarrow}\Big( \hat c_{1\sigma}^\dagger\hat c_{2\sigma}^{\phantom\dagger} + \hat c_{2\sigma}^\dagger\hat c_{3\sigma}^{\phantom\dagger}+\hat c_{1\sigma}^\dagger\hat c_{3\sigma}^{\phantom\dagger}\nonumber\\
&\quad\quad\quad\quad\,\,+\hat c_{2\sigma}^\dagger\hat c_{4\sigma}^{\phantom\dagger}+\hat c_{3\sigma}^\dagger\hat c_{4\sigma}^{\phantom\dagger} + \mathrm{h.c.} \Big)\\
&+2U \sum\limits_{i=1}^4 \hat n_{i\uparrow}\hat n_{i\downarrow}\nonumber,
\label{eq:HSHtHU1}
\end{align}
where $\hat{\vec{S}}_A$ and $\hat{\vec{S}}_B$ represent the spin of the two chromium atoms and $\hat{\vec{S}}_i$ with $i=1,2,3,4$ labeling the spin of the four pyrazine sites. Moreover, $\hat c_{i\sigma}^\dagger$ and $\hat c_{i\sigma}$ are the creation and annihilation operators of an electron on site $i$ and spin $\sigma$, respectively. $\hat n_{i\sigma}=\hat c_{i\sigma}^\dagger\hat c_{i\sigma}\phantom{\dagger}$ is the spin density on site $i$ with spin $\sigma$. $J$ is the exchange coupling and $t$ the hopping constant. 

\begin{figure}
\centering
\includegraphics[width=\linewidth]{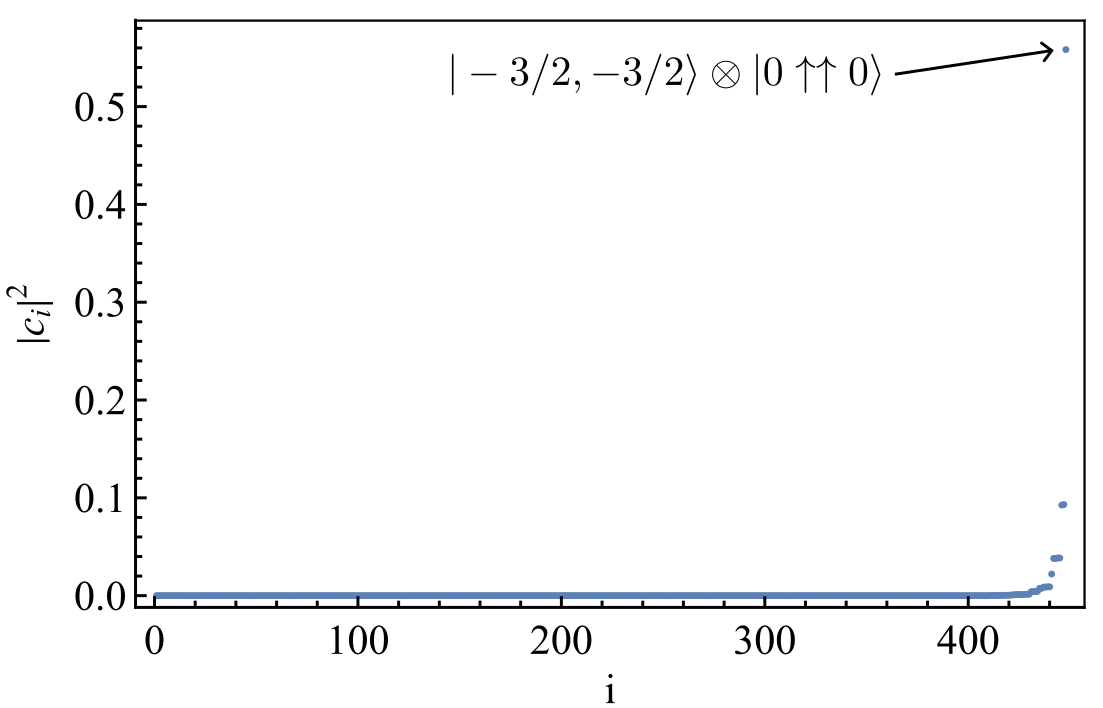}
\caption{A sorted list of the occupation probability of each basis state for one of two degenerate ground states for $t=U=-J=1$. For the illustrated ground state the dominant basis state is $|-3/2,-3/2\rangle \otimes |0\uparrow \uparrow 0\rangle$. It is evident that the probability of all other basis states is small compared to the most dominant basis state. For symmetry reasons the plot of occupation probabilities for the other ground state (when sorted) looks identical, though the dominant basis state here is $|3/2, 3/2\rangle \otimes |0\downarrow\downarrow0\rangle$.}
\label{fig:sojleplot}
\end{figure}

The spins of each chromium atom can take four $S_z$ values, i.e., $-3/2$, $-1/2$, $1/2$ and $3/2$. Details of the diagonalization are provided in the Supplementary Material. The basis is given by $|ms_A, ms_B\rangle \otimes |\psi\rangle$, where $|\psi\rangle$ and $|ms_A, ms_B\rangle$ are the given states for the pyrazine and chromium electrons, respectively.
To invoke an antiferromagnetic coupling between the spin of the pyrazine electrons and the chromium spins, $J$ must be negative. The eigenstates are generally superpositions of 448 states but when all three variables, $J$, $t$ and $U$, are comparable in size the ground state is two-fold degenerate. Surprisingly, however, two basis states in these superpositions turns out to dominate, namely
\begin{equation}
\begin{aligned}
&|-3/2,-3/2\rangle \otimes |0\uparrow\uparrow0\rangle\quad \text{and}\\
&|3/2,3/2\rangle \otimes |0\downarrow\downarrow0\rangle.
\label{eq:v1v2}
\end{aligned}
\end{equation}
This is a robust feature of the model checked for the physically relevant parameter ranges of $t, U$ and $J$. For $t=U=-J=1$ the amplitudes of these states are $-0.747$, and is $\geq 0.5$ for all parameters considered. A two level model is exploited in the Supplementary Material which indicates that one unit cell acts as a spin-2 particle. For $t=U=-J=1$ the occupation probability of the basis states for one of these degenerate states is illustrated, see FIG.~\ref{fig:sojleplot}, clearly showing the dominance of one basis state.

\begin{figure}[h!]
\centering
\includegraphics[width=\linewidth]{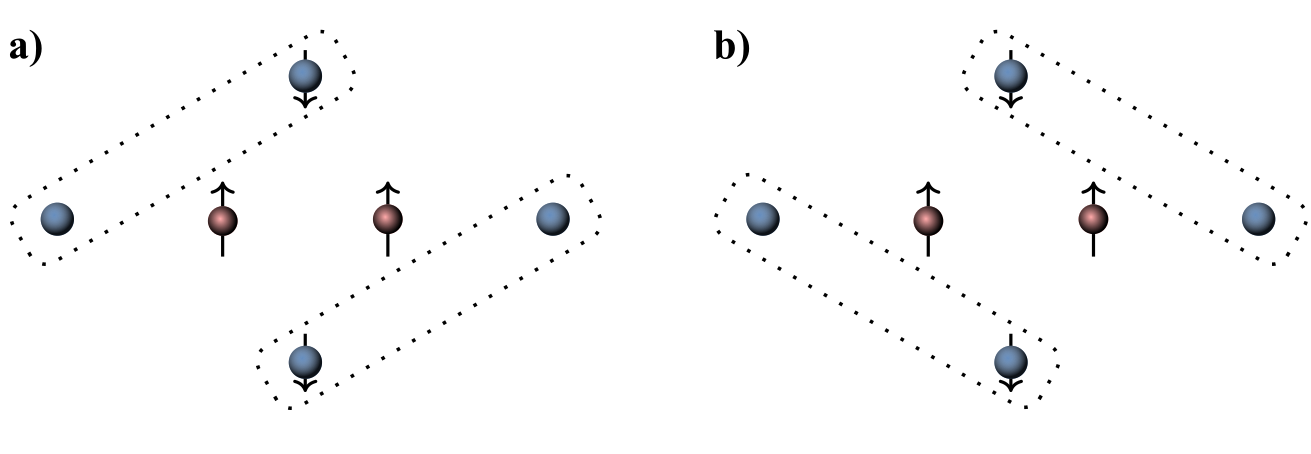}
\caption{The two different configurations of the spins of the pyrazine sites in the minimal model. One electron delocalizes over two sites with different exchange energy and in this way gain kinetic energy while still minimizing the on-site repulsion.}
\label{fig:hoptwoconfig}
\end{figure}

This result has a simple, intuitive explanation. The exchange interaction is maximal when both the pyrazine electrons and the chromium spins are antiparallel and maximal. Electrons located on sites 2 and 3 couple to both chromium spins, unlike electrons located on site 1 and 4 which only couple to one of the Cr spins. Thus, the exchange interaction is further maximized by the electrons being located at site 2 and 3. For $t=0$ this effect is further enhanced and the amplitudes of the states shown in Eq.~\eqref{eq:v1v2} increases to $-0.845$. When $t\neq0$ the electrons can gain energy by delocalizing and the amplitudes of the triplet states with $S_z=0$ and the singlet states increases. The two pyz electrons can delocalize and still avoid each other in two different configurations for a given spin configuration. One electron can delocalize over sites 1 and 2 while the other delocalizes over 3 and 4, see FIG.~\ref{fig:hoptwoconfig} \textbf{a)}, or one electron delocalizes over sites 1 and 3 while the other one delocalizes over sites 2 and 4, see FIG.~\ref{fig:hoptwoconfig} \textbf{b)}.

A magnetic field can be introduced by adding the term 
\begin{align}
\hat{\mathcal{H}}_B=g\mu_B\mathbf{B}\cdot\hat{\mathbf{S}}_{\mathrm{tot}}=g\mu_BB^z\hat S_{\mathrm{tot}}^z,
\label{eq:HB}
\end{align}
to the Hamiltonian. We choose the magnetic field to point out of the plane. 
The expectation value of total spin in the minimal model is found to be $\langle S_{\mathrm{tot}}^z\rangle=\partial E/\partial B=2$. This is derived in the Supplementary Material. Since the model contains the same number of pyrazine rings and chromium atoms as two unit cells of CrCl$_2$(pyz)$_2$ this would correspond to one unit cell having $S=1$ which corresponds to a magnetic moment of
\begin{align}
    \mu_{\text{model}}=2\mu_B.
\end{align}
This is true both for $t=0$ and $t\neq 0$. This is strikingly similar to both experimental results~\cite{Pedersenetal} and DFT results~\cite{Pedersenetal,ChemC}. Experimentally, it is found that the saturation magnetization at high pressure and low temperature is $1.8\mu_B$~\cite{Pedersenetal}. The slightly lower value can be ascribed to the canted radical spins in the real compound and the higher spin occupation on the pyrazines in DFT calculations (0.68 compared 0.5 in this model). Thus, we have here made a minimal model focusing on the relevant physics in the system that is quantitatively precise in its predictions on physical properties.

\begin{figure}[h!]
\centering
\includegraphics[width=\linewidth]{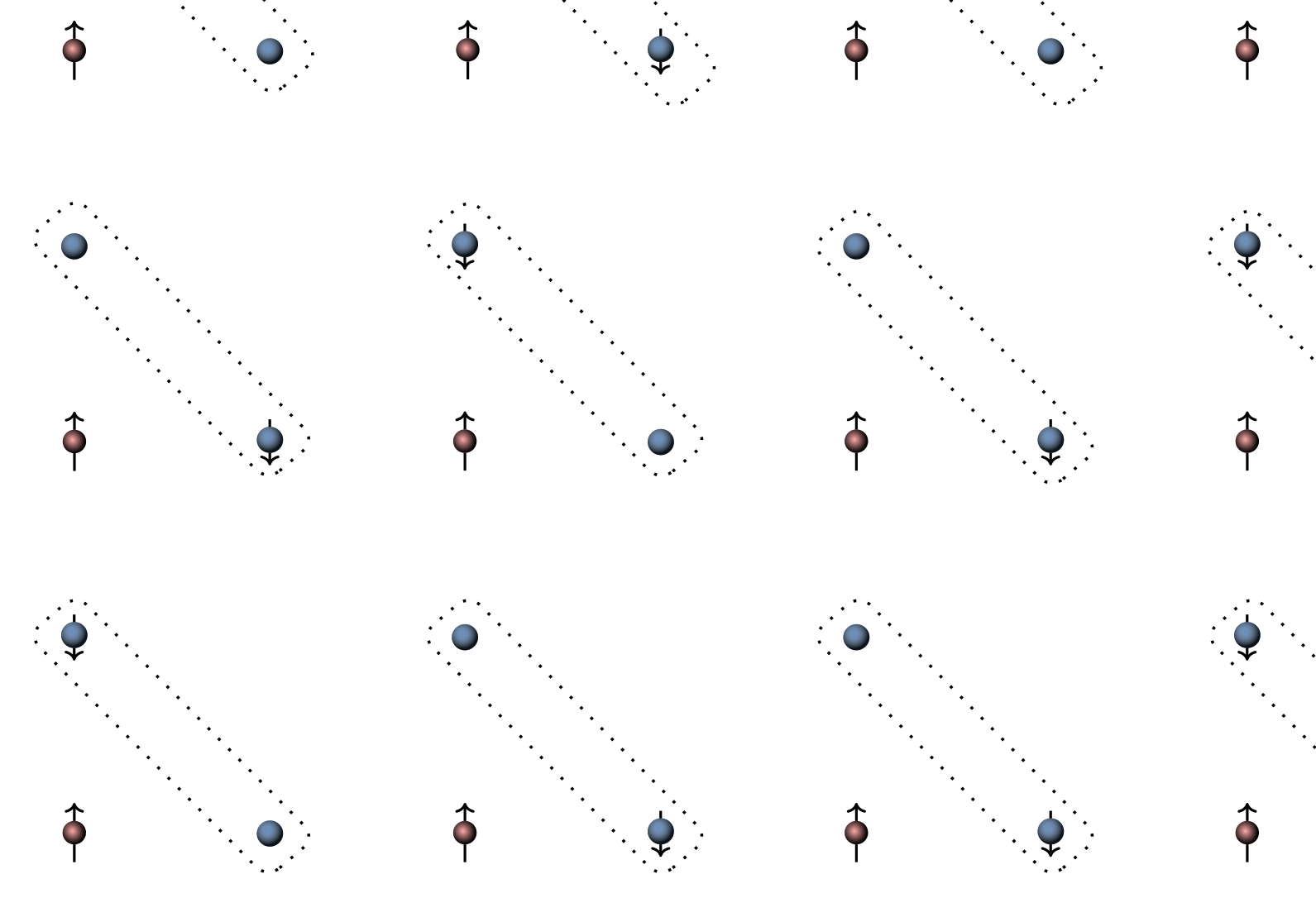}
\caption{A sketch of one of the configurations the delocalized spins on the pyrazine ligands could choose.}
\label{fig:CrCl2pyz2simplemodel}
\end{figure}

If we generalize this model to a layer of CrCl$_2$(pyz)$_2$ we would expect the ground state to look something like what is depicted in FIG.~\ref{fig:CrCl2pyz2simplemodel}. Again, all Cr spins are localized and couple antiferromagnetically to the spins on the pyrazine electrons which have delocalized over two neighboring sites. Obviously, this can be configured in several different ways, but it will focus on one of these, since it will be very costly to change configuration.
\newline

Weiss mean-field theory for a Cr$^{3+}$ square lattice that ignores the pyrazine rings yields a fairly weak ferromagnetic coupling, $J_{\text{Cr-Cr}}=\frac{3k_BT_C}{\mu_0zS(S+1)} \approx 0.9$ meV ($J_{\text{Cr-Cr}}/hc \approx 7$ cm$^{-1}$), see the Supplementary Material for further details. 
A more sophisticated estimate of the Cr-Cr coupling can be obtained by means of second order perturbation theory of the RKKY interaction mediated by intermediate pyrazine rings. This calculation yields $J_{\text{Cr-Cr}} \approx 0.15 J^2/2t$, where $t$ is an effective square lattice hopping strength. Using a strong antiferromagnetic coupling of $J_{\text{Cr-pyz}}\equiv J \approx -0.25$ eV as suggested by ab initio calculations~\cite{Pedersenetal} and $t\approx 1$ eV (see FIG.~\ref{fig:goodstuff}), we predict that $J_{\text{Cr-Cr}} \approx 5$ meV $\left(J_{\text{Cr-Cr}}/hc \approx 40~\text{cm}^{-1}\right)$. Inserting an effective Cr spin of $S=1$, accounting for the presence of the pyz spins, brings the mean-field result closer to the perturbative estimate of the Cr-Cr coupling. 
Here, two independent calculations unify in a meV range prediction of the ferromagnetic Cr-Cr coupling.
This coupling may be probed by neutron scattering techniques and spin wave experiments. Uniaxial strain experiments on CrCl$_2$(pyz)$_2$ may be employed to strain-tune $t$ and reveal potentially interesting insight into the importance of the itinerant electrons and the broken four-fold symmetry. Derivations for both expressions of $J_{\text{Cr-Cr}}$ is included in the Supplementary Material.
\FloatBarrier
\section{Conclusions}
\FloatBarrier
We have proposed both realistic and minimal models capable of explaining the ferrimagnetic state of monolayer CrCl$_2$(pyz)$_2$.
From the Slater--Koster-derived band structure monolayer CrCl$_2$(pyz)$_2$ is found to exhibit an insulating ground state, in agreement with first principle calculations and the experimental results~\cite{Pedersenetal}. Furthermore, the d-bands are observed to be flat indicative of these electrons being localized whereas the pyrazine electrons are found to be very delocalized. 

To investigate the magnetic properties of monolayer CrCl$_2$(pyz)$_2$ a simple model was proposed. Here, the electrons on the chromium sites are viewed as localized spins, whereas the electrons on the pyrazine sites can delocalize over a few sites. This suggests a magnetic moment of one unit cell to be $2 \mu_B$ which corresponds to an antiferromagnetic interaction between the chromium spins and the pyrazine spins. Due to the uneven magnitudes of the spin on the Cr and pyz sites, this results in a ferrimagnetic ordering. The calculated magnetic moment is strikingly close to the experimental value of $1.8\mu_B$~\cite{Pedersenetal}. 
Weiss mean-field theory estimates a direct ferromagnetic coupling between neighboring Cr sites of $J_{\text{Cr-Cr}} \approx 0.9$ meV $\left(J_{\text{Cr-Cr}}/hc \approx 7~\text{cm}^{-1}\right)$. 
Second order perturbation theory of the RKKY interaction between Cr and pyz yields a Cr-Cr coupling of the same order of magnitude, namely $J_{\text{Cr-Cr}}\approx5$ meV. This Cr-Cr coupling may be probed by neutron scattering and spin wave experiments.

\FloatBarrier
\begin{acknowledgments}
FLG acknowledges helpful conversations with Ivano E. Castelli and elaborate discussions and proofreading by Henrik S. Guttesen. 

\end{acknowledgments}

\bibliography{mylib}

\onecolumngrid

\clearpage
\newpage

\setcounter{equation}{0}
\setcounter{figure}{0}
\setcounter{table}{0}
\makeatletter
\renewcommand{\theequation}{S\arabic{equation}}
\renewcommand{\thefigure}{S\arabic{figure}}

\widetext

\begin{center}
\textbf{\large --- Supplementary Material ---}
\end{center}

This is based on results that are previously reported in Ref.~\cite{speciale}. 

\FloatBarrier
\section{Group Theory for Model Complex $\textbf{CrCl}_\textbf{2}\textbf{(pyz)}_\textbf{4}$}
\label{section:supp_grouptheory}
\begin{figure}[h!]
\centering
\includegraphics[width=0.4\linewidth]{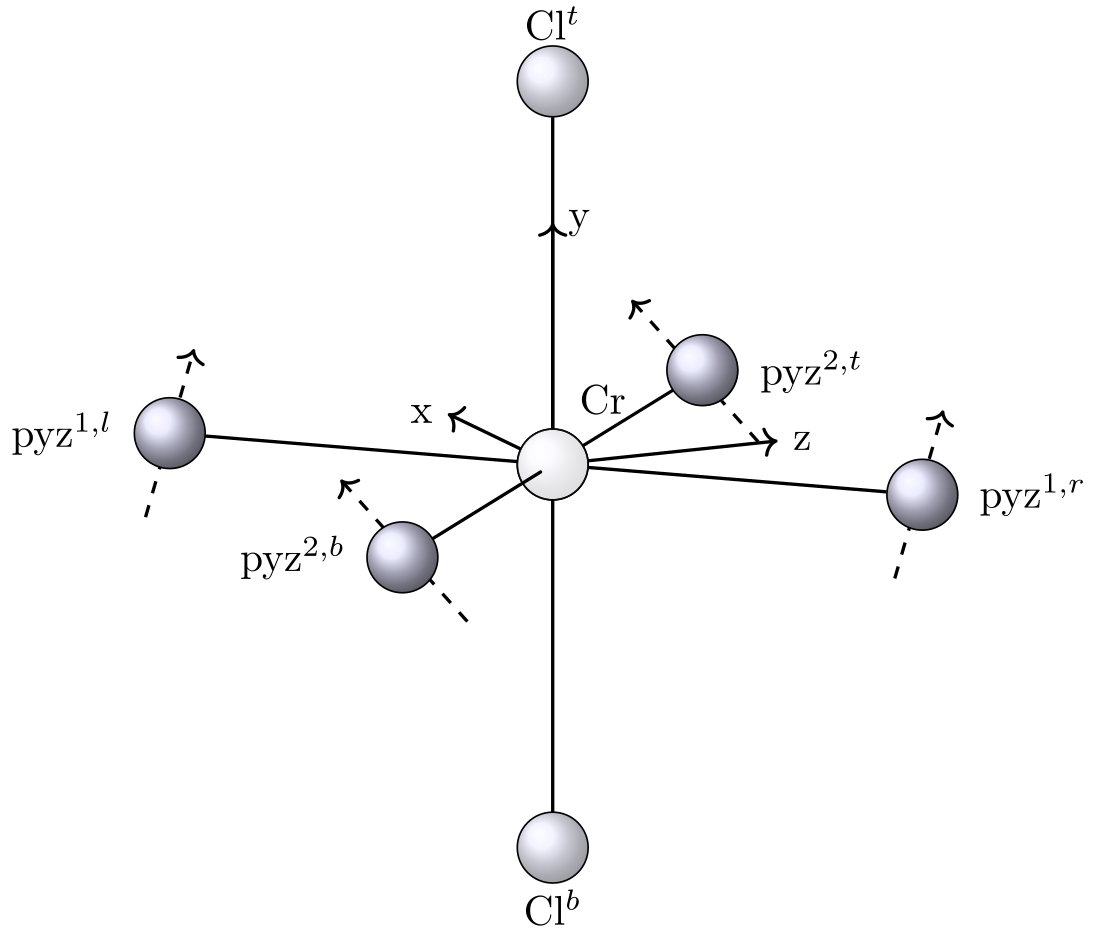}
\caption{Illustration of the model complex CrCl$_2$(pyz)$_4$. The nomenclature of the six relevant atoms, the coordinate system in use, and the orientation of the pyrazine rings are illustrated.
}
\label{fig:onemolCrCl}
\end{figure}

\FloatBarrier
First, we investigate the energy levels and symmetries of the model complex CrCl$_2$(pyz)$_4$, see Fig.~\ref{fig:onemolCrCl}. For this to apply to monolayer CrCl$_2$(pyz)$_2$, the stoichiometry used in this section will be that of CrCl$_2$(pyz)$_2$. One unit cell of CrCl$_2$(pyz)$_2$ consists of one Cr ion, two Cl ions, and two pyrazine rings. In the main text and when investigating monolayer CrCl$_2$(pyz)$_2$ in the following this corresponds to only including pyz$^\text{1,r}$ and pyz$^\text{2,t}$ of the pyz sites in Fig.~\ref{fig:onemolCrCl}. Each chlorine atom reduce the number of electrons on the chromium atom by one. The two pyrazine rings reduce the Cr atom of an extra electron, such that three electrons remain in an unfilled d-shell on the Cr ion. If these spins occupy three distinct energy levels and align parallel to each other, to obey Hund's rules, the spin on the Cr atoms is $S=3/2$.

\begin{figure}[h!]
\centering
\includegraphics[width=.8\linewidth]{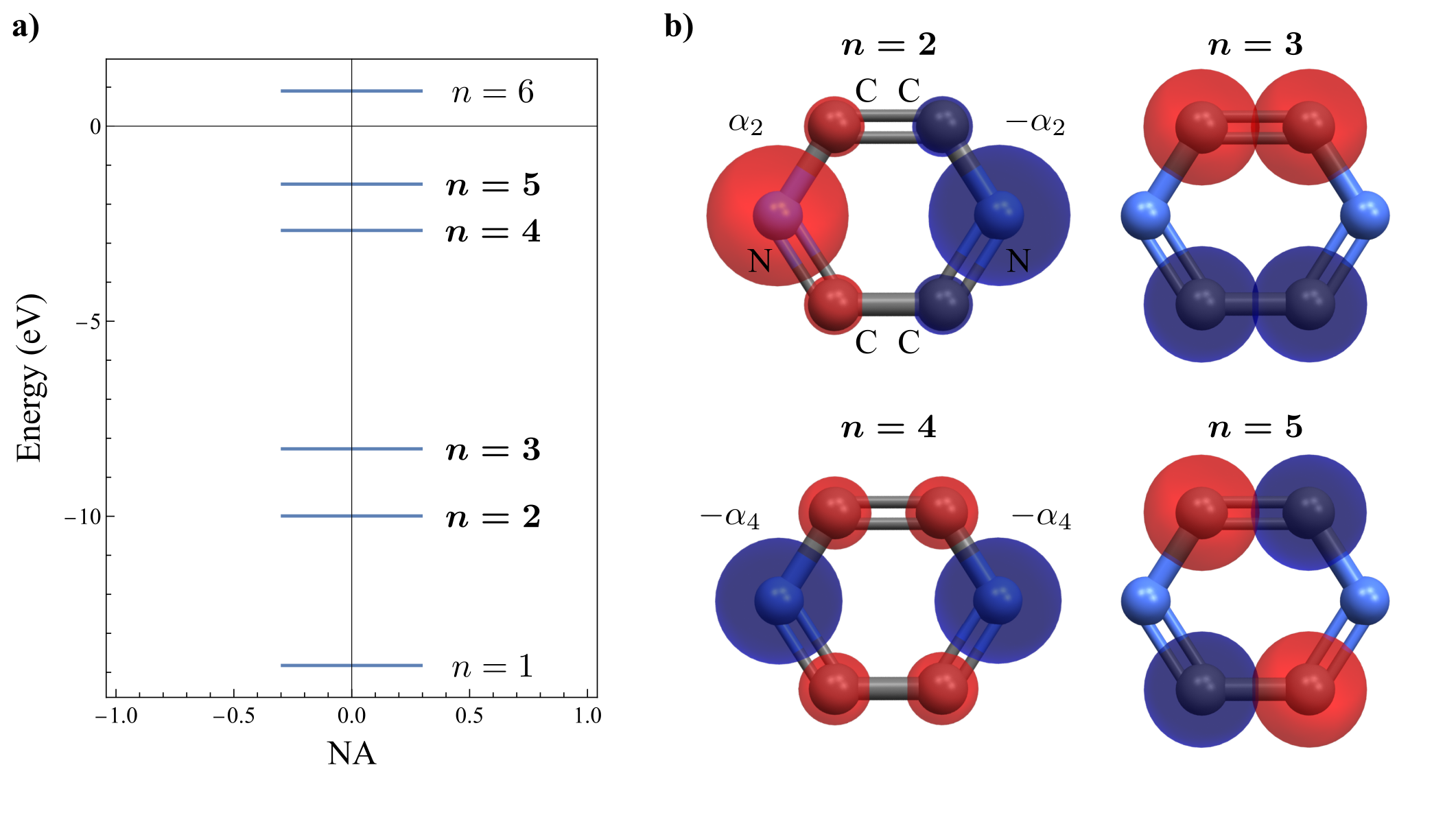}
\caption{\textbf{a)} Energy levels of a pyrazine ring calculated from tight binding. We have excluded the 4 hydrogen atoms and only included the unhybridized p-orbitals on the 4 carbon atoms and 2 nitrogen atoms. \textbf{b)} The eigenstates of the 4 intermediate energy levels $n=2$ to $n=4$. The eigenstates are illustrated by circles with a radius proportional to the amplitude of the orbital occupying the given atom with a positive (red) or negative (blue) sign, respectively.}
\label{fig:pyzlevels}
\end{figure}

As for Benzene, the 6 atoms in the pyrazine rings are sp$^2$ hybridized so we are left with one unhybridized p-orbital per site perpendicular to the plane of the ring. This will be referred to as the $z'$ direction. Due to the presence of the N atoms in pyrazine, compared to Benzene, there are no degenerate states within the ring. The 6 energy levels for a pyrazine ring, where the H atoms are neglected and only the unhybridized p-orbitals are included, is shown in Fig.~\ref{fig:pyzlevels} \textbf{a)}. In Fig.~\ref{fig:pyzlevels} \textbf{b)} the eigenstates for the $n=1,...,4$ levels are illustrated. Each eigenstate is visualized as circles plotted on top of the pyrazine ring with a radius proportional to the amplitude of the p-orbital on the given atom. The red and blue color corresponds to a positive and negative sign on the basis state, respectively.
For the HOMO level ($n=3$) and the LUMO+1 level ($n=5$) the amplitude on the N atoms are equal to zero. These orbitals do not couple to the Cr atoms and will be discarded in the following calculations. The coupling between the carbon atoms in these states and the chromium atom will be negligibly small due to distance. The $n=1$ ($n=6$) level is so high (low) in energy that these can be neglected. Therefore, we only include the two levels $n=2$ and $n=4$, and each pyrazine ring is safely approximated as a single atom with p$_{z'}$-orbitals of two different amplitudes, $\alpha_2$ and $\alpha_4$, where $\alpha_2$ and $\alpha_4$ are the amplitudes on the N atoms of the $n=2$ and $n=4$ levels, respectively. The amplitude on the C atoms are neglected. Note, that the former is an odd linear combination whereas the latter is an even linear combination of the two N atoms.

\FloatBarrier
The pyrazine rings are canted relative to the Cl-Cr-Cl axis. In Ref.~\onlinecite{Pedersenetal} the rotation angles are measured to $42.5^{\degree}$ and $43.9^{\degree}$ with respect to the crystallographic c direction for the pyrazine ring 1 and 2, respectively. This will be approximated as $45^{\degree}$ in both cases. This means that the $p_{z'}$-orbitals of the pyrazine ``atoms'' are instead linear combinations of both $p_x$, $p_y$ and $p_z$ orbitals with respect to the global coordinate system, see FIG.~\ref{fig:onemolCrCl}.

The energy spectrum of model complex CrCl$_2$(pyz)$_4$ can be obtained using the tight binding model, which in second quantization can be written as 
\begin{align}
\hat{\mathcal{H}}_t=\xi\sum\limits_{i, \sigma}\hat c_{i,\sigma}^\dagger\hat c_{i,\sigma}-t\sum\limits_{\langle i,j\rangle}\sum\limits_{\sigma}\left( \hat c_{i,\sigma}^\dagger\hat c_{j,\sigma}+h.c. \right).
\label{eq:tightbinding}
\end{align}
Here $\xi$ refer to the onsite energies, $t$ is the hopping integral and h.c. includes the Hermitian conjugate terms. $\hat c_{i,\sigma}^\dagger$ and $\hat c_{i,\sigma}$ are the creation and annihilation operators, respectively, for an electron on site $i$ with spin $\sigma$. Only nearest neighbor interactions are included. For simplicity the spin-indices will be omitted in the following.

The full Hamiltonian for a single molecule can be written in terms of the Slater--Koster elements in the basis 
\begin{equation}
\begin{aligned}
\Psi=\{ &d_{xy}, d_{yz}, d_{zx}, d_{x^2-y^2},d_{z^2},p_x^t,p_y^t,p_z^t,p_x^b,p_y^b,p_z^b,\\ 
&p^{1,r}_{z'}(\alpha_2),p^{1,r}_{z'}(\alpha_4),p^{2,t}_{z'}(\alpha_2),p^{2,t}_{z'}(\alpha_4),p^{1,l}_{z'}(\alpha_2),p^{1,l}_{z'}(\alpha_4),p^{2,b}_{z'}(\alpha_2),p^{2,b}_{z'}(\alpha_4)\},
\end{aligned}
\end{equation}
where the first five elements refer to the d-orbitals of the chromium atom followed by the three $p$-orbitals of both chlorine atoms (top and bottom), and lastly the p$_{z'}$-orbitals for the four pyrazine ``atoms". The full Hamiltonian can be written in matrix form as
\begin{align}
\mathcal{H}=
\begin{bmatrix}
\mathcal{H}_d & V_{dp} \\
V_{pd} & \mathcal{H}_p
\end{bmatrix}
.
\label{eq:H23x23}
\end{align}
$\mathcal{H}_d$ is a diagonal 5$\cross$5 matrix with $\xi_d$ on the diagonal whereas $\mathcal{H}_p$ is a 14$\cross$14 matrix with $\xi_p$ on the diagonal, where $\xi_d$ and $\xi_p$ are the onsite energies of the d- and p-orbitals, respectively. Thus, the onsite energies of all p-orbitals are assumed to be degenerate prior the interaction with the d-orbitals and similar for the d-orbitals. The two off-diagonal blocks of the Hamiltonian $V_{dp}$ and $V_{pd}$ couple the d-orbitals to the p-orbitals and vice versa.

\FloatBarrier
\subsection{Subspace of the d-orbitals}
\label{section:d-subspace}
The Hamiltonian describing the system can be split up into a part that describes the d-orbitals, a part that describes the p-orbitals and then a term that describes the interaction between these orbitals,
\begin{align}
\hat{\mathcal{H}} = \hat{\mathcal{H}}_p+\hat{\mathcal{H}}_d+\hat{V}.
\end{align}
An eigenstate can be expressed as a sum of a p- and d-part of the state as
\begin{align}
|\psi\rangle = |\psi_p\rangle+|\psi_d\rangle.
\end{align}
The projection operators $\hat{\mathscr{P}}_d$ and $\hat{\mathscr{P}}_p$ for the d-subspace and the p-subspace, respectively, can be applied to rewrite the Hamiltonian. These projection operators must sum to identity, $\hat{\mathscr{P}}_p+\hat{\mathscr{P}}_d=\mathds{1}$, and are defined to project the eigenstate of the full Hamiltonian into the subspace of either the d- or the p-orbitals, as
\begin{align}
\hat{\mathscr{P}}_d|\psi\rangle=E|\psi_d\rangle \quad , \quad \hat{\mathscr{P}}_p|\psi\rangle=E|\psi_p\rangle.
\end{align}
Using these two operators yields
\begin{align}
  \hat{\mathcal{H}}_d |\psi_d\rangle + \hat{V}_{dp}|\psi_p\rangle&=E|\psi_d\rangle,
    \label{eq:1}
    \\
  \hat{\mathcal{H}}_p |\psi_p\rangle + \hat{V}_{pd}|\psi_d\rangle&=E|\psi_p\rangle.
  \label{eq:2}
\end{align}
Here $V_{dp}$ and $V_{pd}$ refer to the 5$\cross$14 and 14$\cross$5 off-diagonal blocks of Eq.~\eqref{eq:H23x23}, respectively. Isolating $|\psi_p\rangle$ in Eq.~\eqref{eq:2} and inserting this in Eq.~\eqref{eq:1} leads to the expression
\begin{align}
\hat{\mathcal{H}}_d|\psi_d\rangle + \hat V_{dp}\frac{1}{E-\hat{\mathcal{H}}_p}\hat V_{pd}|\psi_d\rangle=E|\psi_d\rangle.
\end{align}
Using that $|\psi_d\rangle$ and $|\psi_p\rangle$ are eigenstates of $\hat{\mathcal{H}}_d$ and $\hat{\mathcal{H}}_p$, respectively, and assuming all p-levels to be degenerate prior to the interaction, this can be written as
\begin{align}
\xi_d |\psi_d\rangle+\frac{\xi^2}{E-\xi_p} |\psi_d\rangle = E|\psi_d\rangle.
\end{align}
Here, we have introduced the eigenenergy of $V_{dp}V_{pd}$ as $\xi^2$. Setting $\xi_d=0$, $E$ is isolated as
\begin{align}
E_{\pm} = \frac{\xi_p}{2}\left(1\pm\sqrt{1+\left( \frac{2\xi}{\xi_p}\right)^2}\right). 
\label{eq:Epm}
\end{align}
Since we are seeking the bonding orbitals and $\xi_p$ is chosen to be positive, the $E_+$ solution is discarded. Knowing the eigenenergies of the 5$\cross$5 Hamiltonian $V_{pd}^\dagger V_{pd}$, namely $\xi^2$, in the d-subspace allow us to solve Eq.~\eqref{eq:Epm}. Investigating this matrix analytically we find that the matrix can be written in a block diagonal in the basis $\{ d_{yz}, d_{zx}, d_{xy}, d_{x^2-y^2}, d_{z^2} \}$, 
\begin{align}
V_{dp}V_{pd}=
\begin{bmatrix}
M_{11}&M_{12}&0&0&0\\
M_{12}&M_{22}&0&0&0\\
0&0&M_{11}&M_{34}&M_{35}\\
0&0&M_{34}&M_{44}&M_{45}\\
0&0&M_{35}&M_{45}&M_{55}\\
\end{bmatrix}
,
\label{eq:5x5dCrCl}
\end{align}
with
\begin{align}
M_{11}&=\frac{1}{2}|V_{pd\pi}|^2(4+\alpha),\\
M_{12}&=-\frac{\sqrt{3}}{2}V_{pd\sigma}V_{pd\pi}^\dagger\alpha,\\
M_{22}&=\frac{3}{2}|V_{pd\sigma}|^2\alpha,\\
M_{34}&=\frac{1}{4}V_{pd\pi}^\dagger\alpha\left(\sqrt{2}V_{pd\pi}-\sqrt{3}V_{pd\sigma}\right),\\
M_{35}&=-\frac{1}{4}V_{pd\pi}^\dagger\alpha\left(\sqrt{6}V_{pd\pi}+V_{pd\sigma}\right),\\
M_{45}&=\frac{1}{8}\left[ 4\sqrt{3}|V_{pd\sigma}|^2-\alpha \left( 2\sqrt{3}V_{pd\pi}+\sqrt{2}V_{pd\sigma} \right) \left( V_{pd\pi}^\dagger - \sqrt{\frac{3}{2}}V_{pd\sigma}^\dagger \right) \right],\\
M_{44}&=\frac{1}{8}\left[ 12\sqrt{3}|V_{pd\sigma}|^2+\alpha \left( 2V_{pd\pi}-\sqrt{6}V_{pd\sigma} \right) \left( V_{pd\pi}^\dagger - \sqrt{\frac{3}{2}}V_{pd\sigma}^\dagger \right) \right],\\
M_{55}&=\frac{1}{8}\left[ 4 |V_{pd\sigma}|^2 +\alpha\left( V_{pd\pi}^\dagger \left(6V_{pd\pi}+\sqrt{6}V_{pd\sigma}\right) +V_{pd\sigma}^\dagger (V_{pd\sigma}+\sqrt{6}V_{pd\pi})\right) \right],
\end{align}
where $\alpha\equiv|\alpha_2|^2+|\alpha_4|^2$ has been defined. Here, $V_{ijb}$ is the $b=\sigma$ or $b=\pi$ bond integral between two orbitals $i=\text{(s, p, d, f, ...)}$ and $j=\text{(s, p, d, f, ...)}$. Note that $M_{11}=M_{33}$ by symmetry. The two blocks in Eq.~\eqref{eq:5x5dCrCl} are trivial to diagonalize. It is evident from Eq.~\eqref{eq:5x5dCrCl} that the $d_{yz}$ orbital only couples to the $d_{zx}$ orbital, and vice versa, and further that $d_{xy}$, $d_{x^2-y^2}$, and $d_{z^2}$ only couples to each other. To understand this, we elaborate on the symmetries of the molecule in the following. 

The fact that the pyrazine rings are rotated by two different angles decreases the symmetry of the molecule $C_{\text{4h}} \rightarrow C_{\text{2h}}$. As always, the group describing the spatial symmetries of CrCl$_2$(pyz)$_4$ includes identity. Only three other operations leave the molecule invariant, namely inversion (i), a $\pi$ rotation of about the $z$-axis ($C_2(z)$) and a reflection in the $xy$ plane ($\sigma_{\mathrm{h}}$), see FIG.~\ref{fig:onemolCrCl}. These four symmetry-operations form the point group $C_{2\mathrm{h}}$. This group is Abelian and thus both the order of the group and the number of irreducible representations are equal to 4. The character table can now be constructed, see Table~\ref{table:characterC2h}. 

\begin{table}[h!]
\centering
\caption{Character table of the $C_{2\mathrm{h}}$ group.}
\begin{tabular}{l l c c c c }
    \hline \hline
        $C_{2\mathrm{h}}$ & $E$& $C_2(z)$ & i & $\sigma_{\mathrm{h}}$ & Quadratic Functions \\\hline
        \vspace{-0.25cm}\\ 
    $\mathrm{A}_{\mathrm{g}}$ & 1 & 1 & 1 & 1 & $x^2$, $y^2$, $z^2$, $xy$\\
        $\mathrm{B}_{\mathrm{g}}$ & 1 & -1 & 1 & -1 & $yz$, $zx$ \\
        $\mathrm{A}_{\mathrm{u}}$ & 1&1&-1&-1 &\\
        $\mathrm{B}_{\mathrm{u}}$ & 1&-1&-1&1 &\\ \hline \hline
\end{tabular}
\label{table:characterC2h}
\end{table}
\noindent
From the character table it is observed that $yz$ and $zx$ belong to the same representation and similarly for the last three $d$-orbitals. This explains the block diagonal form of Eq.~\eqref{eq:5x5dCrCl}. Further, $\mel{xy}{\mathcal{H}}{xy} =\mel{yz}{\mathcal{H}}{yz}$ which is due to the chosen coordinate system conforming with the symmetry of the model complex such that these matrix elements are identical.

\FloatBarrier
\section{Tight Binding Model of Monolayer $\textbf{CrCl}_\textbf{2}\textbf{(pyz)}_\textbf{2}$}
\label{sec:supp_TBBandstructure}
\FloatBarrier
Monolayer CrCl$_2$(pyz)$_2$, see FIG.~\ref{fig:goodstuff} \textbf{b)}, is now modeled. Using the Rietveld refinement technique~\cite{Rietveld} the three lattice constants are in Ref.~\cite{Pedersenetal} found to be $a=6.90351(4)$ Å, $b=6.97713(4)$ Å and $c=10.82548(6)$ Å. The close affinity of $a$ and $b$ motivates the lattice to be approximated as a square lattice in the Cr-pyz plane. The relevant symmetry-points are thus approximated to be 
\begin{align}
\mathbf{\Gamma}=(0,0,0), \quad \mathbf{M}=\left(0,0,\frac{1}{2}\right), \quad \mathbf{X}=\left(-\frac{1}{2},0,\frac{1}{2}\right).
\end{align}
These symmetry points are indicated on the Brillouin zone illustrated in FIG.~\ref{fig:goodstuff} \textbf{c)}. 

A unit cell includes one chromium atom, two chlorine atoms and two pyrazine rings, which will be labeled 1 and 2. Assuming that the two pyrazine rings have the same onsite energy, and similarly for the chlorine atoms, a tight binding Hamiltonian for this system in real-space, including only nearest neighbor hopping between p- and d-orbitals, can be written as
\begin{align}
\hat{\mathcal{H}} = & \xi^{\mathrm{Cr}}\sum\limits_{\mathbf{r}_i}\sum\limits_{\mu=1}^5 \hat d_{i,\mu}^\dagger\hat d_{i,\mu}^{\phantom\dagger} + \xi^{\mathrm{Cl}}\sum\limits_{\mathbf{r}_i}\sum\limits_{\nu=1}^3\left[ (\hat p_{i,\nu}^t)^\dagger\hat p_{i,\nu}^t + (\hat p_{i,\nu}^b)^\dagger\hat p_{i,\nu}^b \right]\nonumber\\
& + \xi^{\mathrm{pyz}}\sum\limits_{\mathbf{r}_i} \left[ (\hat p_{i,z'}^{1})^\dagger\hat p_{i,z'}^{1} + (\hat p_{i,z'}^{2})^\dagger\hat p_{i,z'}^{2} \right]\nonumber\\
& + \frac{1}{2}\sum\limits_{\mathbf{r}_i}\sum\limits_{\mathbf{r}_{j_\pm}=\mathbf{r}_i\pm\mathbf{a}_1} \sum\limits_{\mu=1}^5 \left[ t_{\mu,\alpha}^{1,r} \hat d_{i,\mu}^\dagger \hat p_{j_+}^1 + t_{\mu,\alpha}^{1,l} \hat d_{i,\mu}^\dagger \hat p_{j_-}^1 + \mathrm{h.c.} \right]
\label{eq:TB_realspace} \\
& + \frac{1}{2}\sum\limits_{\mathbf{r}_i}\sum\limits_{\mathbf{r}_{j_\pm}=\mathbf{r}_i\pm\mathbf{a}_2} \sum\limits_{\mu=1}^5 \left[ t_{\mu,\alpha}^{2,t} \hat d_{i,\mu}^\dagger \hat p_{j_+}^2 + t_{\mu,\alpha}^{2,b} \hat d_{i,\mu}^\dagger \hat p_{j_-}^2 + \mathrm{h.c.} \right]\nonumber\\
& +\sum\limits_{\mathbf{r}_i}\sum\limits_{\mathbf{r}_{j_\pm}=\mathbf{r}_i\pm\mathbf{a}_3}\sum\limits_{\mu=1}^5 \left[ t_{\mu,\nu}^t \hat d_{i,\mu}^\dagger \hat p_{j_+}^t + t_{\mu,\nu}^b \hat d_{i,\mu}^\dagger \hat p_{j_-}^b + \mathrm{h.c.} \right].\nonumber
\end{align}
Here $\hat{d}_{i,\mu}^\dagger$ ($\hat{d}_{i,\mu}^{\phantom\dagger}$) creates (annihilates) an electron on the chromium site at $\mathbf{r}_i$ in unit cell $i$ and orbital $\mu$. Similarly, the operator $(\hat{p}_{j,\nu}^{t,b})^\dagger$ and $\hat{p}_{j,\nu}^{t,b}$ creates and annihilates an electron in the orbital $\nu$ at the chlorine atom at site $\mathbf{r}_i\pm\mathbf{a}_3$, respectively. Lastly, the operators $(\hat{p}_{j,z'}^{1 (2)})^\dagger$ and $\hat{p}_{j,z'}^{1 (2)}$ creates and annihilates an electron in a $p_{z'}$ orbital on the pyrazine site 1 (2) at $\mathbf{r}_i+\mathbf{a}_1$ and $\mathbf{r}_i+\mathbf{a}_2$, respectively. 

The three first terms of the Hamiltonian states the onsite energies of the d- and p-orbitals of the five atoms in the unit cell. The lower three lines of Eq.~\eqref{eq:TB_realspace} describe electrons hopping from a p-orbital to any nearest neighboring d-orbital. The factor $1/2$ compensate for double counting. For simplicity the spin indices have been omitted. The following on-site energies and hopping integrals are defined as
\begin{align}
\xi^{\mathrm{Cr}}_{e_g}&=\mel{\mathbf{r}_i;d_\mu}{\hat{\mathcal{H}}_t}{\mathbf{r}_i;d_\mu},\\
\xi^{\mathrm{Cl}}&=\mel{\mathbf{r}_i\pm\mathbf{a}_3;p_{\nu}}{\hat{\mathcal{H}}_t}{\mathbf{r}_i\pm\mathbf{a}_3;p_{\nu}},\\
\xi^{\mathrm{pyz}}&=\mel{\mathbf{r}_i+\mathbf{a}_1;p_{z'}(\alpha')}{\hat{\mathcal{H}}_t}{\mathbf{r}_i+\mathbf{a}_1;p_{z'}(\alpha')},\\
t_{\mu,\nu}^t&=\mel{\mathbf{r}_i;d_\mu}{\hat{\mathcal{H}}_t}{\mathbf{r}_i+\mathbf{a}_3;p_\nu},\\
t_{\mu,\nu}^b&=\mel{\mathbf{r}_i;d_\mu}{\hat{\mathcal{H}}_t}{\mathbf{r}_i-\mathbf{a}_3;p_\nu},\\
t_{\mu,z'}^{1,r}&=\mel{\mathbf{r}_i;d_\mu}{\hat{\mathcal{H}}_t}{\mathbf{r}_i+\mathbf{a}_1;p_{z'}(\alpha')},\\
t_{\mu,z'}^{1,l}&=\mel{\mathbf{r}_i;d_\mu}{\hat{\mathcal{H}}_t}{\mathbf{r}_i-\mathbf{a}_1;p_{z'}(\alpha')},\\
t_{\mu,z'}^{2,t}&=\mel{\mathbf{r}_i;d_\mu}{\hat{\mathcal{H}}_t}{\mathbf{r}_i+\mathbf{a}_2;p_{z'}(\alpha')},\\
t_{\mu,z'}^{2,b}&=\mel{\mathbf{r}_i;d_\mu}{\hat{\mathcal{H}}_t}{\mathbf{r}_i-\mathbf{a}_2;p_{z'}(\alpha')}.
\label{eq:ts}
 \end{align}
$\hat{\mathcal{H}}_t$ includes hopping between the two relevant orbitals. Here, we have used that $\alpha'=\alpha_2,\alpha_4$. The numerical values for these can be obtained using the Slater--Koster decomposition~\cite{Harrison,SlaterKoster}. $t^t$ and $t^b$ are 3$\cross$5 matrices that includes hopping between the 3 p-orbitals on the Cl atoms and the 5 d-orbitals on the Cr atoms. The last four lines in Eq.~\eqref{eq:ts} includes the 2$\cross$5 hopping matrices between pyrazine orbitals and the chromium atoms that take the form
\begin{align}
t_{\mu,z'}^{1,r} &=
\begin{bmatrix}
\alpha_2 E_{z',xy}&\alpha_2E_{z',yz}&\alpha_2E_{z',zx}&\alpha_2 E_{z',x^2-y^2}&\alpha_2 E_{z',z^2}\\
\alpha_4 E_{z',xy}&\alpha_4E_{z',yz}&\alpha_4E_{z',zx}&\alpha_4 E_{z',x^2-y^2}&\alpha_4 E_{z',z^2}
\end{bmatrix}_{\mu,z'}, \\
t_{\mu,z'}^{1,l} &=
\begin{bmatrix}
-\alpha_2 E_{z',xy}&-\alpha_2E_{z',yz}&-\alpha_2E_{z',zx}&-\alpha_2 E_{z',x^2-y^2}&-\alpha_2 E_{z',z^2}\\
\alpha_4 E_{z',xy}&\alpha_4E_{z',yz}&\alpha_4E_{z',zx}&\alpha_4 E_{z',x^2-y^2}&\alpha_4 E_{z',z^2}
\end{bmatrix}
_{\mu,z'}.
 \end{align}
The above Slater--Koster elements in the global coordinate system for d-orbital $\mu$ on Cr and the p$_{z'}$-orbitals on pyz is given by 
\begin{align}
E_{z',\mu}=\left[\frac{1}{4}\left( \sqrt{2} -2\right) E_{x,\mu} + 2E_{y,\mu}+\left(2+\sqrt{2}\right) E_{z,\mu}\right].
\end{align}
The bonding matrices $t^{2,t}$ and $t^{2,b}$ take on a similar form. The Hamiltonian can be formulated in reciprocal space as
\begin{align}
\label{eq:Hk}
\hat{\mathcal{H}}=&\sum\limits_{\mathbf{k}} \bigg\{ \xi^{\mathrm{Cr}}\sum\limits_{\mu=1}^5 \hat d_{\mathbf{k},\mu}^\dagger\hat d_{\mathbf{k},\mu}^{\phantom\dagger} + \xi^{\mathrm{Cl}}\sum\limits_{\nu=1}^3 \left[ (\hat p_{\mathbf{k},\nu}^t)^\dagger\hat p_{\mathbf{k},\nu}^t + (\hat p_{\mathbf{k},\nu}^b)^\dagger\hat p_{\mathbf{k},\nu}^b \right]\nonumber\\
&+\xi^{\mathrm{pyz}} \left[ (\hat p_{\mathbf{k},z'}^{1})^\dagger\hat p_{\mathbf{k},z'}^{1} + (\hat p_{\mathbf{k},z'}^{2})^\dagger\hat p_{\mathbf{k},z'}^{2} \right]\nonumber\\
&+ \sum\limits_{\mu=1}^5 \left[ \hat d_{\mathbf{k},\mu}^\dagger\hat p_{\mathbf{k},z'}^{1\phantom{\dagger}}\left(t_{\mu,z'}^{1,r}e^{i\mathbf{k}\cdot\mathbf{a}_1} + t_{\mu,z'}^{1,l}e^{-i\mathbf{k}\cdot\mathbf{a}_1}\right) + \mathrm{h.c.} \right]\\
&+ \sum\limits_{\mu=1}^5 \left[ \hat d_{\mathbf{k},\mu}^\dagger\hat p_{\mathbf{k},z'}^{2\phantom{\dagger}}\left(t_{\mu,z'}^{2,t}e^{i\mathbf{k}\cdot\mathbf{a}_2} + t_{\mu,z'}^{2,b}e^{-i\mathbf{k}\cdot\mathbf{a}_2}\right) + \mathrm{h.c.} \right]\nonumber\\
&+ \sum\limits_{\mu=1}^5\sum\limits_{\nu=1}^3 \left[ t_{\mu,\nu}^t\hat d_{\mathbf{k},\mu}^\dagger\hat p_{\mathbf{k},\nu}^{t\phantom{\dagger}}e^{i\mathbf{k}\cdot\mathbf{a}_3} + t_{\mu,\nu}^b\hat d_{\mathbf{k},\mu}^\dagger\hat p_{\mathbf{k},\nu}^{b\phantom{\dagger}} e^{-i\mathbf{k}\cdot\mathbf{a}_3} + \mathrm{h.c.} \right] \bigg\}\nonumber.
\end{align}

The eigenenergies of the $15\cross 15$ matrix $\hat{\mathcal{H}}$ are obtained numerically for the path $\mathbf{M} \rightarrow \mathbf{\Gamma} \rightarrow \mathbf{X}$ in reciprocal space, yielding the band structure in FIG.~\ref{fig:goodstuff} \textbf{c)}. The following values, obtained from a manual fit to DFT results~\cite{Pedersenetal}, were used to calculate the energy bands, 
$\xi^{\mathrm{Cr}}_{t_{2g}}= -5.7$ eV, $\xi^{\mathrm{Cr}}_{e_{g}}= 1.56$ eV, $\xi^{\mathrm{Cl}}= -6.24$ eV, $\xi^{\mathrm{pyz}}_{\alpha_2}= -1.02$ eV, $\xi^{\mathrm{pyz}}_{\alpha_4}= -0.08$ eV, $V_{pd\sigma}= 1.7$ eV, $V_{pd\pi}= -4.75$ eV, $\alpha_2= 0.67$, $\alpha_4=-0.54$. Knowing that 3 of the d-levels are half-filled and the last two are empty, we employ two different on-site energies for the chromium orbitals; one value for the three $t_{2g}$ orbitals and one for the two $e_g$ orbitals. Further, we know from FIG. \ref{fig:pyzlevels} that the two ``orbitals" on the pyrazine sites have two distinct values. All Cl orbitals are assumed to have the same on-site energy. 

The empty d-bands are located above 1.5 eV (the $e_{g}$ orbitals) whereas the three filled d-bands ($t_{2g}$) are located below the p-levels (not visible in the plot). This is evident when studying the eigenstates of each band but even more clearly seen from first principle calculations of the PDOS~\cite{Pedersenetal}. These d-bands are very flat and thus, the d-electrons are localized on the Cr sites. This could indicate a strong $\pi$-d hybridization. In contrary the p-bands mainly occupied by pyz electrons, located above and below the Fermi energy, are dispersive indicating that the pyz electrons are delocalized.

\FloatBarrier
\section{The Minimal Model and its ground states}
\label{sec:supp_toymodel}
\FloatBarrier
Here, we present an effective model of monolayer CrCl$_2$(pyz)$_2$, see FIG.~\ref{fig:simplemodel}. Note that this is not identical to two unit cells of a single layer of CrCl$_2$(pyz)$_2$ since it contain fewer pyrazine sites and they are shifted with respect to the two chromium atoms. However, it essentially describes the competition between the kinetic energy and the exchange energy. In this model the two chromium atoms are viewed as localized $S=3/2$ spins. There are two electrons that can delocalize over the four pyrazine sites. The chlorine sites are neglected in this model, except for the fact that they have reduced the number of electrons on the chromium sites. 

We include the exchange coupling between the chromium spins and the spins of the two pyrazine electrons. Furthermore, the pyrazine electrons can hop between the pyrazine sites (indicated with dashed lines in FIG.~\ref{fig:simplemodel}). Lastly, we include onsite Hubbard repulsion, $U>0$, for electrons located on the same site. The full Hamiltonian can thus be written as
\begin{equation}
\begin{aligned}
\hat{\mathcal{H}}= &-2J\hat{\vec{S}}_A\cdot\left(\hat{\vec{S}}_1+\hat{\vec{S}}_2+\hat{\vec{S}}_3\right)-2J\hat{\vec{S}}_B\cdot\left(\hat{\vec{S}}_2+\hat{\vec{S}}_3+\hat{\vec{S}}_4\right)\\
&-t\sum\limits_{\sigma=\uparrow,\downarrow}\left( \hat c_{1\sigma}^\dagger\hat c_{2\sigma}^{\phantom\dagger} + \hat c_{2\sigma}^\dagger\hat c_{3\sigma}^{\phantom\dagger}+\hat c_{1\sigma}^\dagger\hat c_{3\sigma}^{\phantom\dagger}+\hat c_{2\sigma}^\dagger\hat c_{4\sigma}^{\phantom\dagger}+\hat c_{3\sigma}^\dagger\hat c_{4\sigma}^{\phantom\dagger} + \mathrm{h.c.} \right)+U \sum\limits_{i=1}^4 \hat n_{i\uparrow}\hat n_{i\downarrow}^{\phantom\dagger},
\label{eq:HSHtHU_supp}
\end{aligned}
\end{equation}
where $\hat{\vec{S}}_A$ and $\hat{\vec{S}}_B$ represent the spin of the two chromium atoms and $\hat{\vec{S}}_i$ with $i=1,2,3,4$ represent the spin of the four pyrazine electrons. The operators $\hat c_{i\sigma}^\dagger$ and $\hat c_{i\sigma}$ create and annihilate an electron on site $i$ with spin $\sigma$, respectively. $J$ is the exchange integral, $t$ the hopping constant and $U$ is the strength of the Hubbard repulsion.

\FloatBarrier
\subsection{Basis}
\FloatBarrier
The spins of each chromium atom can take four $S_z$ values, i.e., $-3/2$, $-1/2$, $1/2$ and $3/2$. The basis for the two electrons on the pyrazine sites will be written in terms of the singlet and triplet states. For this purpose I will introduce the following notation. A singlet state will be represented by a line 
, e.g.,
\begin{align}
&|-00\rangle =\frac{1}{\sqrt{2}} \left( |\uparrow \downarrow 0 0\rangle-|\downarrow \uparrow 0 0\rangle \right),\\
&|\overbow{0}0\rangle=\frac{1}{\sqrt{2}} \left( |\uparrow 0\downarrow 0\rangle-|\downarrow0 \uparrow0\rangle \right),\\
&|-000\rangle =\frac{1}{\sqrt{2}} \left( |\uparrow \downarrow0 0 0\rangle-|\downarrow \uparrow0 0 0\rangle \right).
\end{align}
The first state is a singlet state between an electron on site 1 and an electron on site 2, the second between two electrons on site 1 and 3, and the last is a singlet state of two electrons both occupying site 1. Similarly, we can represent a triplet state with $S_z=0$ as an arrow,
\begin{align}
&|\rightarrow00\rangle =\frac{1}{\sqrt{2}} \left( |\uparrow \downarrow 0 0\rangle+|\downarrow \uparrow 0 0\rangle \right),\\
&|\overarrow{0}0\rangle=\frac{1}{\sqrt{2}} \left( |\uparrow 0\downarrow 0\rangle+|\downarrow0 \uparrow0\rangle \right).
\end{align}
The reason for representing the triplet state as an arrow can be seen by anticommuting operators, $\hat c_{1\uparrow}^\dagger\hat c_{2\downarrow}^\dagger + \hat c_{1\downarrow}^\dagger\hat c_{2\uparrow}^\dagger =- \left(\hat c_{2\uparrow}^\dagger\hat c_{1\downarrow}^\dagger + \hat c_{2\downarrow}^\dagger\hat c_{1\uparrow}^\dagger\right)$, such that the overall sign is meaningfully giving the direction of the arrow. For singlets the two states would be identical. All triplet states are defined such that the operator to the left creates an electron on a site labeled by a lower numerical value than the operator to the right (also applicable to $S_z=1$ and $S_z=-1$ triplets). The basis for the two electrons are thus 
\begin{align}
|\psi\rangle=\Big\{ &|-00\rangle, |0-0\rangle, |00-\rangle, |\overbow{0}0\rangle, |0\overbow{0}\rangle, |\overbow{00}\rangle,\nonumber\\
&|-000\rangle, |0-00\rangle, |00-0\rangle, |000-\rangle,\nonumber\\
& |\downarrow \downarrow00\rangle, |0\downarrow \downarrow0\rangle, |00\downarrow \downarrow\rangle, |\downarrow0 \downarrow0\rangle, |0\downarrow 0\downarrow\rangle, |\downarrow00 \downarrow\rangle,\\
& |\rightarrow00\rangle, |0\rightarrow0\rangle, |00\rightarrow\rangle, |\overarrow{0}0\rangle, |0\overarrow{0}\rangle, |\overarrow{00}\rangle,\nonumber\\
& |\uparrow \uparrow00\rangle, |0\uparrow \uparrow0\rangle, |00\uparrow \uparrow\rangle, |\uparrow0 \uparrow0\rangle, |0\uparrow 0\uparrow\rangle, |\uparrow00 \uparrow\rangle\Big\}.\nonumber
\end{align}
The full state is then given by $\ket{ms_A} \otimes \ket{ms_B} \otimes \ket{\psi}$ and the Hilbert space is 448-dimensional. 
\FloatBarrier
\subsection{Hamiltonian representation}
\FloatBarrier
The Hamiltonian is diagonalized numerically. First, we rewrite the Hamiltonian, using $\hat{\vec{S}}_1\cdot\hat{\vec{S}}_2=\hat S_1^z\hat S_2^z + \frac{1}{2}\left( \hat S_1^+\hat S_2^-+\hat S_1^-\hat S_2^+ \right)$, such that
\begin{equation}
\begin{aligned}
\hat{\mathcal{H}}= &-2J\hat S_A^z\left(\hat S_1^z+\hat S_2^z+\hat S_3^z\right)-J\hat S_A^+\left(\hat S_1^-+\hat S_2^-+\hat S_3^-\right) -J\hat S_A^-\left(\hat S_1^++\hat S_2^++\hat S_3^+\right)\\
&-2J\hat S_B^z\left(\hat S_2^z+\hat S_3^z+\hat S_4^z\right)-J\hat S_B^+\left(\hat S_2^-+\hat S_3^-+\hat S_4^-\right) -J\hat S_B^-\left(\hat S_2^++\hat S_3^++\hat S_4^+\right)\\
&-t\sum\limits_{\sigma=\uparrow,\downarrow}\left( \hat c_{1\sigma}^\dagger\hat c_{2\sigma}^{\phantom\dagger} + \hat c_{2\sigma}^\dagger\hat c_{3\sigma}^{\phantom\dagger}+\hat c_{1\sigma}^\dagger\hat c_{3\sigma}^{\phantom\dagger}+\hat c_{2\sigma}^\dagger\hat c_{4\sigma}^{\phantom\dagger}+\hat c_{3\sigma}^\dagger\hat c_{4\sigma}^{\phantom\dagger} + \mathrm{h.c.} \right)\\
&+U \sum\limits_{i=1}^4\sum\limits_{\sigma,\sigma'=\uparrow,\downarrow} \hat c_{i\sigma'}^\dagger\hat c_{i\sigma}^\dagger\hat c_{i\sigma}^{\phantom\dagger}\hat c_{i\sigma'}^{\phantom\dagger}.
\end{aligned}
\end{equation}
In the following we will explore how the full Hamiltonian can be written in the basis described above by investigating some specific examples. The less enthusiastic reader can potentially skip this part and jump to section~\ref{supp:twolevel}. 

We focus on the first three exchange terms including $\hat S_A^z$. The chromium part of this is trivial in the basis $\{\ket{-3/2}, \ket{-1/2}, \ket{1/2}, \ket{3/2} \}$: 
\begin{align}
M_{\mathrm{Cr}}^z=
\begin{bmatrix}
-\frac{3}{2}&0&0&0\\
0&-\frac{1}{2}&0&0\\
0&0&\frac{1}{2}&0\\
0&0&0&\frac{3}{2}
\end{bmatrix}.
\end{align}
Now, let us see how this affects the basis states for the electrons on the pyrazine sites. Applying this operator on basis states where neither of the two electrons are located on site 4 results in the expected $S_z=0$ for a singlet state, e.g.,
\begin{align}
\left(\hat S_1^z+\hat S_2^z+\hat S_3^z\right) \ket{-00}
 = \frac{1}{\sqrt{2}} \left( \frac{1}{2}\ket{\uparrow\downarrow00}-\frac{1}{2}\ket{\uparrow\downarrow00}-\frac{1}{2}\ket{\downarrow\uparrow00}+\frac{1}{2}\ket{\downarrow\uparrow00} \right) = 0.
\end{align}
The same applies to triplet states with $S_z=0$ with no electrons occupying site 4 and for singlet states where both electrons occupy the same site. On the other hand, when one of the electrons is occupying site 4, singlet and triplet states couple. For instance:
\begin{align}
\left(\hat S_1^z+\hat S_2^z+\hat S_3^z\right) \ket{00-} = \frac{1}{\sqrt{2}}\left[ \frac{1}{2}\ket{00\uparrow\downarrow}-\left(-\frac{1}{2}\ket{00\downarrow\uparrow}\right) \right] =\frac{1}{2} \ket{00\rightarrow}.
\end{align}
The triplet states with $S_z=1$ or $S_z=-1$ are eigenstates of this operator; the eigenvalue depends on whether one of the electrons is occupying site 4 or not,
\begin{align}
&\left(\hat S_1^z+\hat S_2^z+\hat S_3^z\right) \ket{0\downarrow\downarrow0}=-\ket{0\downarrow\downarrow0},\\
&\left(\hat S_1^z+\hat S_2^z+\hat S_3^z\right) \ket{00\downarrow\downarrow}=-\frac{1}{2}\ket{00\downarrow\downarrow},\\
&\left(\hat S_1^z+\hat S_2^z+\hat S_3^z\right) \ket{0\uparrow\uparrow0}=\ket{0\uparrow\uparrow0},\\
&\left(\hat S_1^z+\hat S_2^z+\hat S_3^z\right) \ket{00\uparrow\uparrow}=\frac{1}{2}\ket{00\uparrow\uparrow}.
\end{align}
This procedure is similar for the $S_B^z$ part of the Hamiltonian but in this case singlet and triplet states are coupled when one of the electrons is occupying site 1.

Now, let us look at the other part of the exchange interaction, i.e., the terms\\ $-J\hat S_A^+\left(\hat S_1^-+\hat S_2^-+\hat S_3^-\right) -J\hat S_A^-\left(\hat S_1^++\hat S_2^++\hat S_3^+\right)$. The raising and lowering operators for the chromium spins can be written as
\begin{align}
M_{\mathrm{Cr}}^+=
\begin{bmatrix}
0&0&0&0\\
\sqrt{3}&0&0&0\\
0&2&0&0\\
0&0&\sqrt{3}&0
\end{bmatrix}
\quad\text{and}\quad
M_{\mathrm{Cr}}^-=
\begin{bmatrix}
0&\sqrt{3}&0&0\\
0&0&2&0\\
0&0&0&\sqrt{3}\\
0&0&0&0
\end{bmatrix}.
\end{align} 
For the electrons on the pyrazine site it is not as simple. As before, the outcome depends on whether one of the electrons is occupying site 4 or not. Focusing on the electron part including raising operators for the pyrazine electrons, we look at a few cases to see how the basis states are affected by this operator,
\begin{align}
&\left(\hat S_1^++\hat S_2^++\hat S_3^+\right) \ket{-00}=0,\\
&\left(\hat S_1^++\hat S_2^++\hat S_3^+\right) \ket{00-}=-\frac{1}{\sqrt{2}}\ket{00\uparrow\uparrow}, \\
&\left(\hat S_1^++\hat S_2^++\hat S_3^+\right) \ket{\rightarrow00}=\sqrt{2}\ket{\uparrow\uparrow00}, \\
&\left(\hat S_1^++\hat S_2^++\hat S_3^+\right) \ket{00\rightarrow}=\frac{1}{\sqrt{2}}\ket{00\uparrow\uparrow}, \\
&\left(\hat S_1^++\hat S_2^++\hat S_3^+\right) \ket{00\downarrow\downarrow}=\ket{00\uparrow\downarrow}\nonumber,\\
&\quad\quad=\frac{1}{2}\left( \ket{00\uparrow\downarrow}+\ket{00\downarrow\uparrow} \right)+\frac{1}{2}\left( \ket{00\uparrow\downarrow}-\ket{00\downarrow\uparrow} \right)\nonumber\\
&\quad\quad=\frac{1}{\sqrt{2}}\ket{00\rightarrow}+\frac{1}{\sqrt{2}}\ket{00-}.
\end{align}
The final example is quite interesting since this results in a superposition between a triplet state and a singlet state. The same method can be used to obtain the resulting states when applying the operator $\left(\hat S_1^-+\hat S_2^-+\hat S_3^-\right)$ and similarly for the B-part of the exchange coupling. 
\\
Now, we look at the tight binding part of the Hamiltonian. Since the spin is unchanged by this part, the singlet and the triplet states will not couple, and thus we can look at these parts of the basis states separately. First, let us focus on the 10 singlet states. Here, one example is sufficient to illustrate what happens to all of them:
\begin{align}
\hat{\mathcal{H}}_t\ket{-00}=-t\left( \sqrt{2}\ket{-000}+\sqrt{2}\ket{0-00}+|\overbow{0}0\rangle+\ket{0-0}+|\overbow{00}\rangle \right).
\end{align} 
What is important to note here is that a factor of $\sqrt{2}$ arises when a singly occupied state become a doubly occupied site, since
\begin{align}
\sum\limits_\sigma\hat c_{1\sigma}^\dagger\hat c_{2\sigma} \ket{-00}&=\sum\limits_\sigma\hat c_{1\sigma}^\dagger\hat c_{2\sigma} \frac{1}{\sqrt{2}}\left( \hat c_{1\uparrow}^\dagger\hat c_{2\downarrow}^\dagger- \hat c_{1\downarrow}^\dagger\hat c_{2\uparrow}^\dagger\right)\ket{\mathrm{vac}}\nonumber\\
&=\frac{1}{\sqrt{2}}\left( \hat c_{1\uparrow}^\dagger\hat c_{1\downarrow}^\dagger - \hat c_{1\downarrow}^\dagger\hat c_{1\uparrow}^\dagger \right)\ket{\mathrm{vac}} = \sqrt{2}\ket{-000}.
\end{align}
Here $\ket{\mathrm{vac}}$ represents the vacuum state. Due to the antisymmetry of a singlet state we do not need to think about signs. Again, one example is sufficient:
\begin{align}
\hat{\mathcal{H}}_t\ket{\uparrow\uparrow00}&=\sum\limits_\sigma \left( \hat c_{3\sigma}^\dagger\hat c_{1\sigma} + \hat c_{3\sigma}^\dagger\hat c_{2\sigma} +\hat c_{4\sigma}^\dagger\hat c_{2\sigma}\right)\ket{\uparrow\uparrow00}\nonumber\\
&=\left( \hat c_{3\uparrow}^\dagger\hat c_{1\uparrow}\hat c_{1\uparrow}^\dagger\hat c_{2\uparrow}^\dagger + \hat c_{3\uparrow}^\dagger\hat c_{2\uparrow}\hat c_{1\uparrow}^\dagger\hat c_{2\uparrow}^\dagger +\hat c_{4\uparrow}^\dagger\hat c_{2\uparrow}\hat c_{1\uparrow}^\dagger\hat c_{2\uparrow}^\dagger\right)\ket{\mathrm{vac}}\nonumber\\
&= \left( -\hat c_{2\uparrow}^\dagger\hat c_{3\uparrow}^\dagger +\hat c_{1\uparrow}^\dagger\hat c_{3\uparrow}^\dagger +\hat c_{1\uparrow}^\dagger\hat c_{4\uparrow}^\dagger \right)\ket{\mathrm{vac}}\nonumber\\
&=-\ket{0\uparrow\uparrow0}+\ket{\uparrow0\uparrow0}+\ket{\uparrow00\uparrow}.
\end{align}
The last term in the Hamiltonian (onsite repulsion) is trivial to include for the four singlet states. Using this basis the $448\cross 448$ Hamiltonian can be diagonalized numerically yielding the results described in the main text. The main conclusion when diagonalizing this Hamiltonian is that the two degenerate ground state are dominated by the two basis states viewed in Eq.~\eqref{eq:v1v2} and Fig.~\ref{fig:hoptwoconfig}.

\FloatBarrier
\subsection{Two Level Model}
\label{supp:twolevel}
\FloatBarrier
Focusing on the dominant basis state in the ground state obtained in the model described above, see Eq.~\eqref{eq:v1v2} and Fig.~\ref{fig:hoptwoconfig}, i.e., going to the limit $J\gg t$, we can make a simple model to understand how the ground state energy changes when increasing the hopping parameter $t$. For $t=0$ the two electrons will localize on sites 2 and 3 with spins antiparallel compared to the chromium spins. When $t$ increases the two electrons can then hop between two different sites and still avoid each other, which they can do in two different configurations. One electron can hop between site 1 and 2 while the other hop between 3 and 4, see FIG.~\ref{fig:hoptwoconfig} \textbf{a)}, or one electron can hop between site 1 and 3 while the other can hop between 2 and 4, see FIG.~\ref{fig:hoptwoconfig} \textbf{b)}. 

\begin{figure}[h]
\centering
\includegraphics[width=.5\textwidth]{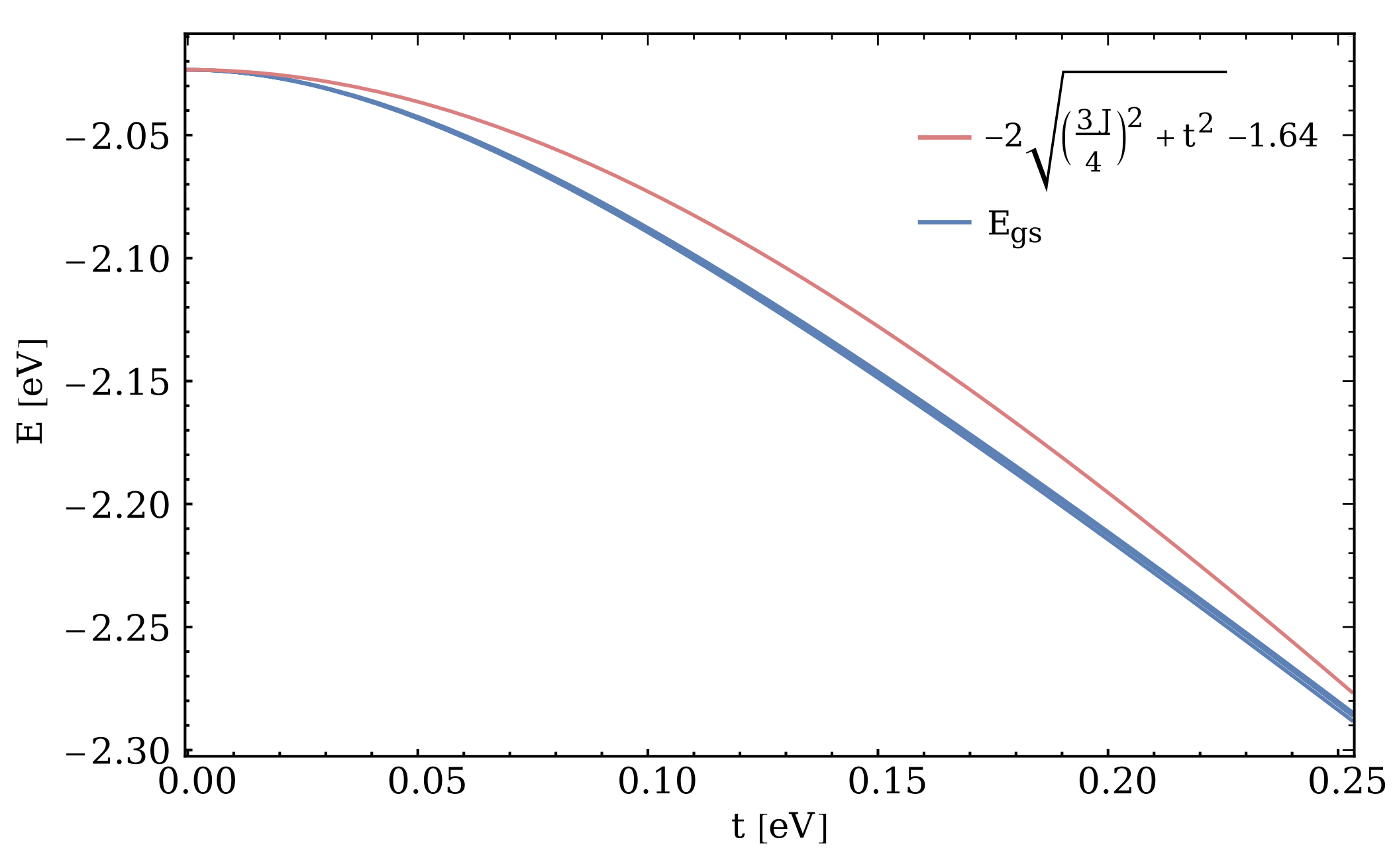}
\caption{The five lowest, degenerate eigenenergies plotted as a function of t with $J=-0.2529$ eV and $U=0.2529$ eV, compared with the function in Eq.~\eqref{eq:eigEJt}, shifted to match at t = 0.}
\label{fig:E_af_t}
\end{figure}

For an electron hopping between two sites, e.g.,~1 and 2, the two levels do not have the same energy, since the electron can couple to both Cr spins at site 2 but only couple to one Cr spin at site 1. The gap between these two levels is $\Delta=\frac{3J}{4}$ since one level has the energy $\left(\frac{3}{2}+\frac{3}{2}\right)J=3J$ and the other has energy $\frac{3}{2}J$. Thus, we employ the simple Hamiltonian 
\begin{align}
\hat{\mathcal{H}}=
\begin{bmatrix}
\frac{3J}{4}&t\\
t&\frac{-3J}{4}
\end{bmatrix},
\end{align}
with the eigenvalues
\begin{align}
E=\pm\sqrt{\left(\frac{3J}{4} \right)^2+t^2}.
\label{eq:eigEJt}
\end{align}
The negative solution is the ground state. In FIG.~\ref{fig:E_af_t} we show this function alongside the exact eigenvalues as a function of $t$ with $J=-0.2529$ eV and $U=0.2529$ eV. The value chosen for $J$ is motivated by Ref.~\cite{Pedersenetal}, with $U$ chosen to be of the same magnitude as $J$. Note that Eq.~\eqref{eq:eigEJt} has been shifted to fit with the exact eigenvalues in the figure. The exact ground state energies follow this simple model quite closely in the entire $t$ regime investigated. The exact solution is a bit lower in energy, which can be explained by the fact that in this model the system has two possible configurations with the same low energy, see FIG.~\ref{fig:hoptwoconfig}. This only lowers the energy slightly since it has to be coordinated such that the electrons can still avoid each other. These features are robust towards changes in $U$ (even turning off repulsion completely) since electrons occupying the same site is not favorable in this model.

When $t=0$ the ground state energy becomes five-fold degenerate. Setting $t\neq0$ this energy splits in three of which two are two-fold degenerate. The five-fold degeneracy in the limit where $J\gg t$ indicates that the system behaves as a spin-2 particle. This makes sense since 
\begin{equation}
\begin{aligned}
\hat S_{\mathrm{tot}}^z\ket{ 3/2}\otimes\ket{3/2}\otimes\ket{0\downarrow\downarrow} &= \left(\hat S_A^z+\hat S_B^z+\hat S_2^z+\hat S_3^z\right)\ket{3/2}\otimes\ket{3/2}\otimes\ket{0\downarrow\downarrow}\\
&=\left(\frac{3}{2}+\frac{3}{2}-\frac{1}{2}-\frac{1}{2}\right)\ket{3/2}\otimes\ket{3/2}\otimes\ket{0\downarrow\downarrow}\\
&=2\ket{3/2}\otimes\ket{3/2}\otimes\ket{0\downarrow\downarrow}.
\end{aligned}
\end{equation}
All these five lowest states are plotted in FIG.~\ref{fig:E_af_t}.

\FloatBarrier
\subsection{Zeeman splitting}
\begin{figure}[h]
\centering
\includegraphics[width=.8\textwidth]{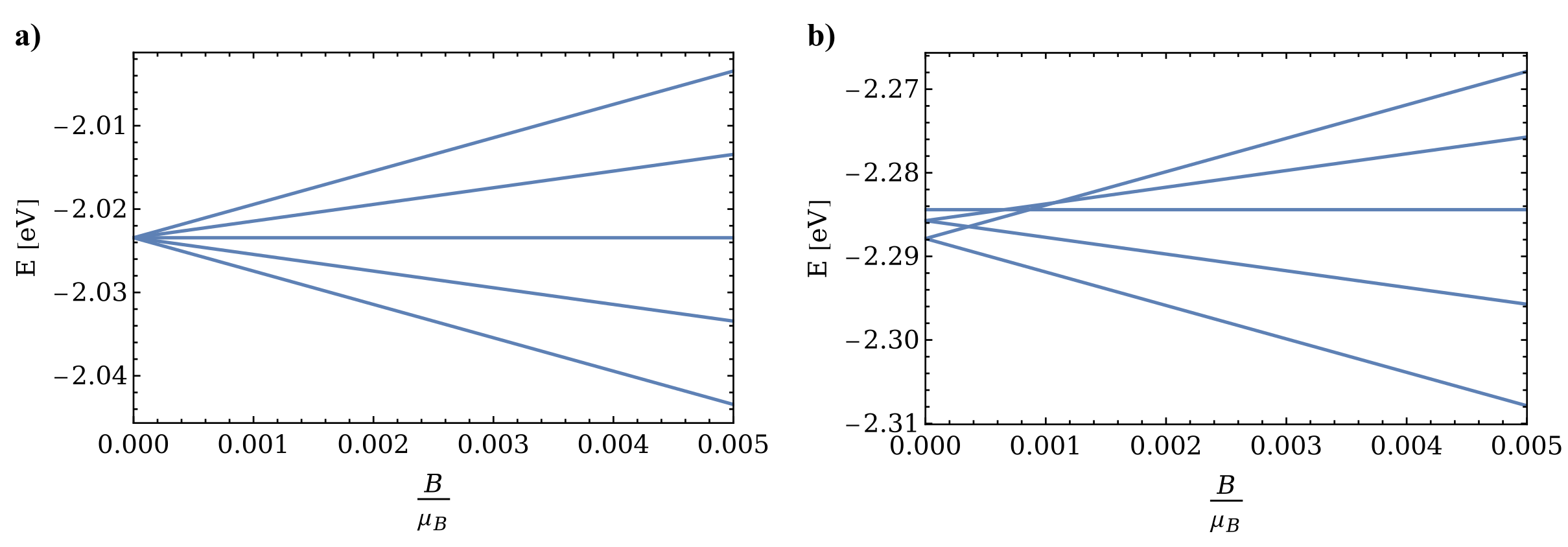}
\caption{The five lowest eigenenergies of Eq.~\eqref{eq:HSHtHU_supp} + Eq.~\eqref{eq:HB_supp} depicted as a function of $B_z$ with $J=-0.2529$ eV$=-U$ for $t=0$ eV (a) and $t=0.2529$ eV (b). The degeneracy is lifted by Zeeman splitting for $B\neq0$.}
\label{fig:E_af_B}
\end{figure}

\FloatBarrier
To see how these states split up in a magnetic field we add the term 
\begin{align}
\hat{\mathcal{H}}_B=g\mu_B\mathbf{B}\cdot\hat{\mathbf{S}}_{\mathrm{tot}}=g\mu_BB^z\hat S_{\mathrm{tot}}^z,
\label{eq:HB_supp}
\end{align}
where we choose the magnetic field to point in the $z$-direction, i.e. out of the plane in FIG.~\ref{fig:simplemodel}. As seen in FIG.~\ref{fig:E_af_B} \textbf{a)} the five-fold degeneracy of the ground state for $t=0$ is lifted when a magnetic field is applied. Furthermore, we see that for $t\neq0$ the energy splits into three, two of which are two-fold degenerate. The two doubly degenerate states act as a doublet and the non-degenerate state behaves as a singlet. 

Adding a magnetic field also allows us to calculate the expectation value of the total spin using
\begin{align}
\expval{S_{\mathrm{tot}}} = \frac{\partial E}{\partial B}.
\end{align} 
Both for $t=0$ and $t\neq0$ this is found numerically to be 2.

\FloatBarrier
\section{Estimate of the effective chromium coupling from mean-field theory}
\label{sec:supp_Curie}
\FloatBarrier

From mean-field theory we can derive an expression of the exchange coupling constant between neighboring chromium spins $\mathbf{S}_i$. We depart from 
\begin{align}
\hat{\mathcal{H}}=-J_{\text{Cr-Cr}}\sum\limits_{\expval{i,j}} \mathbf{S}_i\cdot\mathbf{S}_j.
\end{align}
Using mean-field theory and neglecting the constant shift, the interaction of the $i$th spin is 
\begin{align}
\hat{\mathcal{H}}_i=-J_{\text{Cr-Cr}}z\expval{\mathbf{S}}\cdot \mathbf{S}_i,
\end{align}
where $z$ denotes the number of neighboring spins (the coordination number), i.e., $z=4$ for a square lattice. Defining 
\begin{align}
\mathbf{H}\equiv-\frac{zJ_{\text{Cr-Cr}}}{g\mu_B}\expval{\mathbf{S}},
\end{align}
the Hamiltonian for the $i$th spin can be written as 
\begin{align}
\hat{\mathcal{H}}_i=g\mu_B\mathbf{S}_i\cdot\mathbf{H}.
\end{align}
Thus, $\mathbf{H}$ can be viewed as an exchange field affecting the $i$th spin, which arises from the polarization of the spins on the neighboring chromium sites. The magnetization from this exchange field is then given by 
\begin{align}
\mathbf{M}=-g\mu_BN\expval{\mathbf{S}},
\end{align}
where $N$ is the number of magnetic chromium atoms in the sample. From this an expression for $\expval{\mathbf{S}}$ can be obtained, such that 
\begin{align}
\mathbf{H}=A\mathbf{M}, \quad \text{with} \quad A=\frac{zJ_{\text{Cr-Cr}}}{g^2\mu_B^2N}.
\end{align}
Including an external magnetic field $\mathbf{B}$ the total magnetization magnitude can be written as
\begin{align}
M=\frac{C}{T}\left( B+AM \right),
\end{align}
where $C$ is the Curie constant, which for $L=0$, where $L$ is the total orbital angular momentum, is given by 
\begin{align}
C=\frac{\mu_0\mu_B^2}{3k_B}Ng^2S(S+1).
\end{align}
The magnetization magnitude can now be isolated:
\begin{align}
M=\frac{C}{T-AC}B.
\end{align}
Using the Curie Weiss law, we can conclude that $AC=T_C$, where $T_C$ is the Curie temperature, from which we obtain the expression
\begin{align}
J_{\text{Cr-Cr}}=\frac{3k_BT_C}{\mu_0zS(S+1)}.
\end{align}

\FloatBarrier
\section{Exchange Interaction between $\textbf{Cr}$ and $\textbf{pyz}$}
\label{sec:supp_exchange}
\begin{figure}[h!tb]
\centering
\includegraphics[width=.7\linewidth]{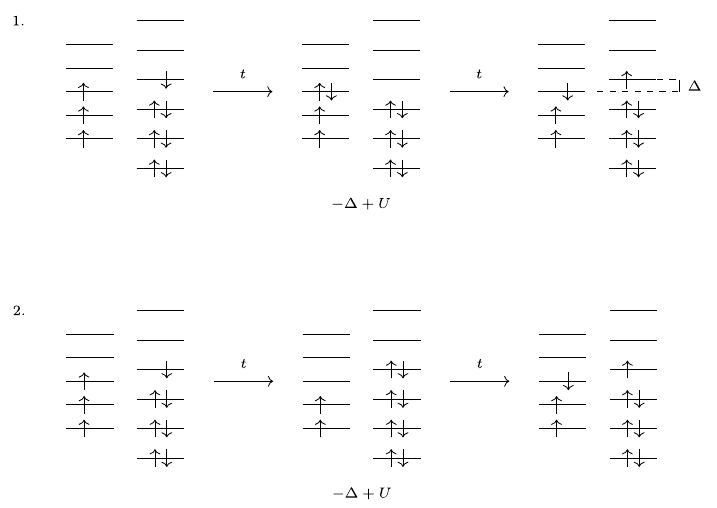}
\caption{Diagram illustrating the two ways in which an electron on the chromium site of one unit cell (five left energy levels in each illustration) can exchange location with an electron on a pyrazine site (6 right energy levels).}
\label{fig:exchangeJCr}
\end{figure}

\FloatBarrier
Here we study the interaction between chromium spins and neighboring pyrazine spins to provide an explanation as to why CrCl$_2$(pyz)$_2$ is ferrimagnetic. An expression for the change in energy, that arise from the interaction, is obtained through second order perturbation theory~\cite{BruusOgFlensberg},
\begin{align}
\Delta E^{(2)}=-\frac{|\langle\psi|\hat{\mathcal{H}}_t|\phi\rangle|^2}{-\Delta+U} - \frac{|\langle\psi|\hat{\mathcal{H}}_t|\phi\rangle|^2}{\Delta}=-|\langle\psi|\hat{\mathcal{H}}_t|\phi\rangle|^2\left( \frac{1}{(1-\Delta/U)\Delta} \right)\equiv J.
\label{eq:JCrpyz}
\end{align}
$|\psi\rangle$ represents the highest occupied energy level on the chromium site in the environment. This is expressed as a superposition of the five d-orbitals and is obtained from the band structure studied previously. $|\phi\rangle$ can be written as 
\begin{align}
|\phi\rangle=\frac{1}{\sqrt{2}}\left( |\phi_1\rangle + |\phi_2\rangle \right),
\end{align}
where $|\phi_1\rangle$ and $|\phi_2\rangle$ are the eigenstates of the relevant energy level on the two pyrazine sites in the unit cell, also obtained from the band structure. We assume that $J/t\ll 1$. $\Delta$ denotes the difference in energy of the two relevant energy levels and $U$ is the strength of the Hubbard repulsion for two electrons occupying the same chromium site, see FIG.~\ref{fig:exchangeJCr}. The Hubbard repulsion for the pyrazine sites is neglected due to their much larger spatial extent compared to chromium. 

In the exchange interaction two electrons exchange location with one intermediate step. This can be done in two ways, both are illustrated in FIG.~\ref{fig:exchangeJCr}. The exchange interaction constant $J$ is a sum over these two possibilities, which results in Eq.~\eqref{eq:JCrpyz}. For the coupling between the Cr sites and the pyrazine sites to be antiferromagnetic, it is required that $\Delta >U$.

\FloatBarrier
\subsection{RKKY-Interaction}
\label{sec:supp_RKKY}
\FloatBarrier
In this section, we derive an expression for the effective chromium coupling, as determined by means of second order perturbation theory of the RKKY interaction mediated through intermediate pyrazine sites.
The starting point of this is the exchange interaction between the spin of the Cr ions, $\mathbf{S}$, and the spin of the valence electron, $\mathbf{\hat s}$, on the pyrazine sites, described above, with the interaction Hamiltonian
\begin{align}
\hat{\mathcal{H}}'=-J\sum\limits_{i,\delta} \mathbf{S}_i \cdot \mathbf{\hat s}_{i+\delta},
\end{align}
for unit cell $i$. $\delta$ sums over the vectors pointing to the four neighboring pyrazine sites. We treat this interaction as a perturbation to the ground state configuration. The second order correction to the energy is given by 
\begin{align}
E_\mathbf{k}^{(2)}=\sum\limits_{\mathbf{k}'\neq \mathbf{k}}\frac{|\langle\psi_{\mathbf{k}'}^0|\hat{\mathcal{H}}'|\psi_\mathbf{k}^0\rangle|^2}{\xi_\mathbf{k}^0-\xi_{\mathbf{k}'}^0}=\sum\limits_{\mathbf{k}'\neq \mathbf{k}}\frac{\langle\psi_{\mathbf{k}}^0|\hat{\mathcal{H}}'|\psi_{\mathbf{k}'}^0\rangle\langle\psi_{\mathbf{k}'}^0|\hat{\mathcal{H}}'|\psi_\mathbf{k}^0\rangle}{\xi_\mathbf{k}^0-\xi_{\mathbf{k}'}^0},
\end{align}
where $\ket{\psi_\mathbf{k}^0}$ is the ground state of the delocalized electrons, i.e., Bloch waves, and where $\xi_{\boldsymbol{k}}^0$ is the normal-state, unperturbed dispersion relation. The chromium spins in $\hat{\mathcal{H}}'$ can be taken outside of the averages since these are not considered as a part of the perturbation. Now, we are left with a product of two correlation functions. Recall that we can write the spin of the valence electrons as 
\begin{align}
\sum\limits_{i}\mathbf{\hat s}_{i+\delta}=\frac{1}{2}\hat c_{i+\delta,\sigma}^\dagger\mathbf{\tau}_{\sigma\sigma'}\hat c_{i+\delta\sigma'}=\frac{1}{2N}\sum\limits_{\mathbf{kk}'}\hat c_{\mathbf{k}\sigma}\mathbf{\tau}_{\sigma\sigma'}\hat c_{\mathbf{k}\sigma'}.
\end{align}
In the textbook version of the RKKY model the valence electrons are considered as a free electron gas. In this case we consider the electrons to be delocalized but anchored to the four pyz sites. For the spin operators in each correlation function we must sum over all bonds to the relevant chromium spin. The dominant terms here are thus the terms where the neighboring chromium atoms couple through an electron located at a pyz site in between. Thus, we find
\begin{align}
E_n^{(2)}=J_{\text{Cr-Cr}}=\frac{J^2}{2N^2}\mathbf{S}_1\cdot\mathbf{S}_2 \,\chi,
\end{align}
where $\chi$ is the Lindhard function given by
\begin{align}
\chi=\,\,
\begin{tikzpicture}[baseline=(current  bounding  box.center)]
\begin{feynman}
  \vertex (a);
  \vertex [right=of a] (b);
  \vertex [right=of b] (c);
  \vertex [right=of c] (d);
  \diagram* {
    (a) -- [photon] (b)
    -- [fermion, half right, looseness=1.5] (c)
    -- [fermion, half right, looseness=1.5] (b), (c) -- [photon] (d),
  };
\end{feynman}
\end{tikzpicture}
\,\,=\sum\limits_{\mathbf{kk}'} \frac{n_F(\xi_\mathbf{k})-n_F(\xi_{\mathbf{k}'})}{\xi_\mathbf{k}-\xi_{\mathbf{k}'}}.
\label{eq:Lindhard}
\end{align}
The factor of $1/N^2$ stems from the Fourier transformation and the factor of two stems from the fact that 
\begin{align}
\langle \psi_{\mathbf{k}\sigma}|\mathbf{\tau}^\alpha\mathbf{\tau}^\beta|\psi_{\mathbf{k}\sigma}\rangle=\mathrm{Tr}[\mathbf{\tau}^\alpha\mathbf{\tau}^\beta]=2\delta_{\alpha\beta}.
\end{align}
At $T=0$ the Fermi functions in Eq.~\eqref{eq:Lindhard} are step functions. The Lindhard function can then be written as 
\begin{align}
\chi=2\sum\limits_{\mathbf{k}>\mathbf{k}_F}\sum\limits_{\mathbf{k}'<\mathbf{k}_F} \frac{1}{\xi_{\mathbf{k}}-\xi_{\mathbf{k}'}}=2\int\limits_{\xi_F}d \xi\int\limits^{\xi_F}d \xi' \frac{D(\xi)D(\xi')}{\xi-\xi'},
\end{align}
where $D(\xi)$ is the density of states. The density of states for a square lattice with spectrum $\xi=2t(\cos k_x+\cos k_y)$ is given by 
\begin{align}
D(\xi)=\frac{N^2}{2(\pi t)^2}K\left[ 1-\left(\frac{\xi}{4t} \right)^2 \right].
\end{align}
Here $K$ is the complete elliptic integral of the first kind. In the subspace of the pyrazine sites there is $1/2$ electron in each unit cell and thus the band has filling $1/4$ due to spin.

The density of states is normalized such that $\int\limits_{-4t}^{4t}d\xi D(\xi)=1.$ Thus, the Fermi energy can be found by requiring that $\int\limits_{-4t}^{\xi_F}d\xi D(\xi)=\frac{1}{4}.$
The Fermi energy is obtained numerically and the Lindhard function is found to have a numerical value of $0.15$ for $t=1$, such that 
\begin{align}
J_{\text{Cr-Cr}}=0.15\frac{J^2}{2t}.
\end{align}
The value $\xi_F/t = 0.15$ obtained within the above approximations demonstrates that the Fermi surface is well-approximated by a small circle.

\end{document}